\documentclass[journal]{IEEEtran}
\usepackage{cite}

\ifCLASSINFOpdf
  \usepackage[pdftex]{graphicx}
  \graphicspath{{../pdf/}{../jpeg/}}
  \DeclareGraphicsExtensions{.pdf,.jpeg,.png}
\else
\fi
\usepackage{amsmath, amssymb, amsfonts}
\usepackage{algorithmic}

\usepackage{array}

\ifCLASSOPTIONcompsoc
  \usepackage[caption=false,font=normalsize,labelfont=sf,textfont=sf]{subfig}
\else
  \usepackage[caption=false,font=footnotesize]{subfig}
\fi
\usepackage{fixltx2e}

\usepackage[caption=false]{subfig}
\usepackage{xcolor}

\graphicspath{{./figures/}}

\begin{document}
%
\title{Rip Current Detection in Nearshore Areas through UAV Video Analysis with Almost Local-Isometric Embedding Techniques on Sphere}
%
%
%

\author{Anchen~Sun,~\IEEEmembership{Student Member,~IEEE,}, 
Kaiqi Yang
        
\thanks{A. Sun is with the Department
of Electrical and Computer Engineering, University of Miami, Coral Gables,
FL, USA. E-mail: asun at miami dot edu. K. Yang is with the Department
of Mathematics, University of Miami, Coral Gables,
FL, USA. E-mail: kxy1105 at miami dot edu.}
}

%
%

\markboth{Journal of \LaTeX\ Class Files,~Vol.~14, No.~8, August~2015}%
{Shell \MakeLowercase{\textit{et al.}}: Bare Demo of IEEEtran.cls for IEEE Journals}
%



\maketitle

\begin{abstract}
Rip currents pose a significant danger to those who visit beaches, as they can swiftly pull swimmers away from shore. Detecting these currents currently relies on costly equipment and is challenging to implement on a larger scale. The advent of unmanned aerial vehicles (UAVs) and camera technology, however, has made monitoring near-shore regions more accessible and scalable. This paper proposes a new framework for detecting rip currents using video-based methods that leverage optical flow estimation, offshore direction calculation, earth camera projection with almost local-isometric embedding on the sphere, and temporal data fusion techniques. Through the analysis of videos from multiple beaches, including Palm Beach, Haulover, Ocean Reef Park, and South Beach, as well as YouTube footage, we demonstrate the efficacy of our approach, which aligns with human experts' annotations.
\end{abstract}


%
\IEEEpeerreviewmaketitle

\section{Introduction}
%
%
%
%
\IEEEPARstart{R}{ip} currents pose a significant threat to beachgoers, and they are responsible for a majority of surf beach lifeguard rescues and drowning incidents~\cite{commerce_noaa_2020}. Since most beaches lack lifeguard protection, they can be especially dangerous during rip current events. Although lifeguards are the most effective means of preventing drowning caused by rip currents, they are costly, resulting in most US beaches being unguarded~\cite{branche2001lifeguard}. To tackle this issue, it is crucial to develop a low-cost, automated device that can accurately estimate wave speed and rapidly detect rip currents in real time. This would potentially reduce the number of drowning cases resulting from these currents. This research paper proposes such a device and evaluates its effectiveness in detecting rip currents.

For decades, coastal scientists and oceanographers have been exploring rip current modeling. In the early 21st century, the analytic model was introduced, which employs the linear stability characteristics of time-averaged flow in rip currents~\cite{haller2001rip}. However, these models require a costly set of equipment to gather data. For instance, the analytic rip current model relies on ten capacitance wave gauges and three acoustic Doppler velocimeters (ADVs) for data collection. The collection of data in such energetic rip current environments necessitates trained and experienced operators, and their equipment and operators are susceptible to risks. Several remote sensing techniques, such as marine radars that rely on the backscattered signal from the rough ocean surface, have been proposed to estimate current velocity and structure~\cite{haller2014rip}. During the past two decades, video-based methods have also been developed~\cite{holman2013remote}. Several studies have focused on one-dimensional measures~\cite{chickadel2003optical}, two-dimensional measures and cross-correlation techniques~\cite{holland2001quantification}, as well as other tracking methods~\cite{kennedy2004drifter}. This research paper discusses various rip current detection methods, including video-based techniques, and compares their effectiveness.

The real-time and dense estimation of rip current speed in nearshore areas is challenging with current techniques. However, recent advancements in machine learning and multimedia data analysis have demonstrated the effective and efficient analysis of unstructured data, including images and videos~\cite{ota2017deep}. As a result, it is possible to automatically detect and estimate the rip current speed using videos captured by an Unmanned Aerial Vehicle (UAV). Compared to conventional model-based rip current speed estimation, video-based estimation using UAV has several distinct advantages, such as:
\begin{enumerate}
	\item UAV video-based approach is adaptable and can be employed in various locations and oceanographic conditions, making it a versatile tool for coastal monitoring.
	\item UAV monitoring coverage is less expensive, and the required equipment is readily available and more straightforward to use than traditional methods.
	\item Model-based algorithms typically require a significant amount of initial data to account for various terms in mathematical and physical equations, making them slower compared to other methods.
\end{enumerate}

Recently, the use of optical flow estimation method has been introduced for analyzing motion and estimating rip current speed based on nearshore UAV videos~\cite{typhoon}. This method has several advantages, as previously mentioned. In \cite{typhoon}, a wavelet-based optical flow estimation method was proposed to estimate the fluid speed at a fine-grained level. The experimental results demonstrated that their method was effective in estimating the speed with an acceptable bias. However, the proposed optical flow estimation method primarily focuses on instant surface currents. As far as the authors are aware, there is no existing framework for automatically detecting rip currents based on videos directly.

This research paper proposes a novel rip current detection framework based on unmanned aerial vehicle (UAV) videos, which detects rip current regions using only video data with the UAV fly logs. The paper makes the following contributions:
\begin{enumerate}
	\item The first completely automated framework for detecting rip currents based on UAV videos.
    \item A new modeling approach, known as almost local-isometric embedding on a sphere, is introduced to precisely estimate the captured pixel on the actual Earth.
	\item A technique is introduced for fusing temporal data from UAV videos to improve the robustness of rip current detection in identified regions.
	\item Analyzing the performance of different optical flow estimation methods and identifying key factors to consider when selecting an appropriate method for current velocity estimation and rip current detection.
	\item Experiments are conducted using videos collected from two different locations to evaluate the performance of our rip current detection framework both qualitatively and quantitatively. The results demonstrate that our proposed method can effectively detect the regions of rip currents.
\end{enumerate}

The structure of this paper is organized as follows. Section~\ref{sec:related} briefly introduces two types of techniques to estimate current speed and detect rip currents, namely model-based and video-based methods. In Section~\ref{sec:method}, we present the details of our proposed framework for UAV-video-based rip current detection, including the temporal data fusion technique, the almost local-isometric embedding modeling on the sphere, and the optical flow estimation methods. Section~\ref{sec:experiment} evaluates the proposed framework using two videos collected from different locations, and provides qualitative and quantitative analyses. Finally, the conclusion and future work are discussed in Section~\ref{sec:conclusion}.

\section{Related Work}\label{sec:related}
\subsection{Model-Based Rip Current Detection Methods}
A variety of methodologies have been used by researchers to identify rip currents, with the most common approach being direct observation with the naked eye. There are several tell-tale signs of rip currents, including a tongue of sediment-laden water moving offshore, seaward movement of floating objects, unusual wave choppiness, foamy water in the outer edges of the rip head, gap in the breaking waves, and darker water that indicates the presence of a rip channel.

One study, referenced as~\cite{sonu1972field}, used water-filled polyethylene balls that were adjusted to be neutrally buoyant to detect rip currents. Comparisons with dye releases indicated that the movement of the freely-drifting balls represented the mean flow. Additionally, nearly-filled plastic jugs of water have been used to locate rips and provide an indication of Lagrangian trajectories, which is a low-cost method that is still widely used in developing countries~\cite{inman1980field}. This method can also provide an estimate of rip speed.

In recent years, more sophisticated instrumentation has been deployed at rip-prone beaches, such as tripod-mounted current meters and pressure sensors. However, there are challenges associated with this Eulerian method of rip measurement, including concerns about instrumentation placement and whether the current meter remains in the mean flow area of the rip. Additionally, logistical problems and safety constraints exist when installing instrumentation in a dangerous location~\cite{brander2000morphodynamics}.

In Australia, experiments involving individuals floating freely in rip currents have been conducted using theodolites~\cite{short1994rip} and more recently GPS technology. Lagrangian methods have gained popularity for measuring rip currents, with GPS-equipped drogues offering advantages such as shorter set-up time, safer deployment, increased mobility, and the ability to track the rip current path. MacMahan et al. extensively used GPS-controlled drogues to measure rip current velocity in California, France, and Australia, with real-time differential GPS drifters having an accuracy of less than 1 meter after carrier phase post-processing~\cite{schmidt2003gps}. Non-differential GPS units, which are significantly less expensive, are now being used with great success to study rip currents~\cite{sabet2011design}.

Smith and Laergier employed Doppler sonar to measure rip currents at the Scripps Institute of Oceanography pier in California~\cite{smith1995observations}. While acoustic meters provide accurate measurements~\cite{chickadel2003optical}, they are expensive, unwieldy, and require significant set-up time, making them unsuitable for rapid deployments or long-term monitoring. Lushine developed the first rip forecast models, which were later refined by Lascody~\cite{lascody1998east}. Lushine's model quantified the rip threat by analyzing drowning, meteorological, and oceanographic records, and established the relationship between rip occurrence and wind and low tide. Lascody modified the model by incorporating swell wave data~\cite{lascody1998east}.

Numerical models that predict rip currents require detailed bathymetric and offshore boundary condition data. Several coupled 3-D circulation and wave propagation models have been developed to predict rip currents. Kumar et al.\cite{kumar2011implementation} coupled the circulation model ROMS with the wave propagation model SWAN to model rip currents. ROMS is a 3-D numerical bathymetry-following circulation model that solves finite difference approximations of Reynolds-averaged Navier-Stokes equations using hydrostatic and Boussinesq approximations with a split-explicit time stepping algorithm\cite{kumar2011implementation}. Brown et al.\cite{brown2009surf} applied the Delft 3-D circulation model to the nearshore with promising results. Wang et al.\cite{wang2018numerical} used the FUNWAVE model, which utilizes the Boussinesq equations, to simulate rip currents off arc-shaped coastlines. These numerical models require significant amounts of initial data and processing time and are not suitable for real-time measurements of rip currents.

Video cameras have been a popular tool for measuring waves and currents at beaches for many years~\cite{sonu1972field}. These cameras are typically mounted on buildings, poles, or balloons and use telephoto lenses to capture footage of the surf zone. Holman and Stanley used an Argus camera system at the Corps of Engineers Field Research Facility in Duck, North Carolina to obtain time-lapsed photography~\cite{holman2007history}. Google Earth and Bing satellite images can also be used to identify rip currents based on differential water coloration or sediment plumes. Optical methods, such as Particle Image Velocimetry (PIV), have been used to study rip currents~\cite{holland2001quantification, chickadel2003optical}. These techniques use video footage to track the drift of sea foam in the surf zone and derive current velocity measurements. However, they require a set of visible ground control points with known locations and are limited by the location and placement of the camera, as well as the contrast of features in the images. Chickadel et al. developed an optical current meter algorithm that Fourier transforms space-time data into a frequency-wave number spectrum and then to a velocity spectrum~\cite{chickadel2003optical}.

\subsection{Video-Based Optical Flow Estimation Methods}

A number of studies have explored the use of video cameras to detect and measure rip currents. Gallop et al. conducted a pilot study to estimate the responses of rip channels to wave energy and wave event duration using a conceptual model based on 40 months of rip channel data obtained from automated analysis of video imagery~\cite{gallop2009video}. Other studies have focused on observing rip current velocity using video recordings of floating tubes placed in the sea at Haeundae Beach, Korea~\cite{bae2013boussinesq, yoon2014observation}. To improve the accuracy of rip current measurements, Stresser et al. recorded videos using a self-stabilizing camera gimbal mounted on a small quadcopter~\cite{stresser2017video}. They used the resulting image sequences to determine spatio-temporal characteristic parameters of the short-term surface wave, and fitted a linear dispersion relationship to the data to monitor the frequency shift caused by the currents. However, while these studies provide valuable insights into the characteristics of rip current features extracted from videos, they were limited to coarse-level detection of rip currents.

The estimation of dense fluid velocity from video has been a topic of interest since the 20th century. Optical flow estimation methods have emerged as an effective approach to address this problem~\cite{horn1981determining, lucas-k}. In \cite{kadri2013divergence}, the integration of wavelets and optical flow estimation methods was proposed for the instant estimation of surface currents from shore-based and UAV videos. Derian et al. developed the "Typhoon" algorithm to extract the apparent displacements (motion) from image sequences specifically for fluid velocity estimation~\cite{derian2015wavelet}. The algorithm estimates two-dimensional fluid velocity fields from fluid-flow sensory data, such as particle image velocimetry (PIV), schlieren photography, or lidar imagery. The ``Typhoon" algorithm was further extended and utilized to estimate the instant surface currents from shore-based and UAV videos in 2017~\cite{typhoon}. The velocity field estimated by the ``Typhoon" algorithm was validated by comparison with measurements from an acoustic Doppler velocimeter, demonstrating its ability to capture both wave-to-wave fluctuations and low-frequency variations. The method was also applied successfully to monitor a ``flash rip" event. This research has clearly demonstrated the potential of optical flow estimation methods for near-shore video analysis.

\section{Proposed Rip Current Detection Framework}\label{sec:method}
\begin{figure}
	\centering
	\includegraphics[width=0.35\textwidth]{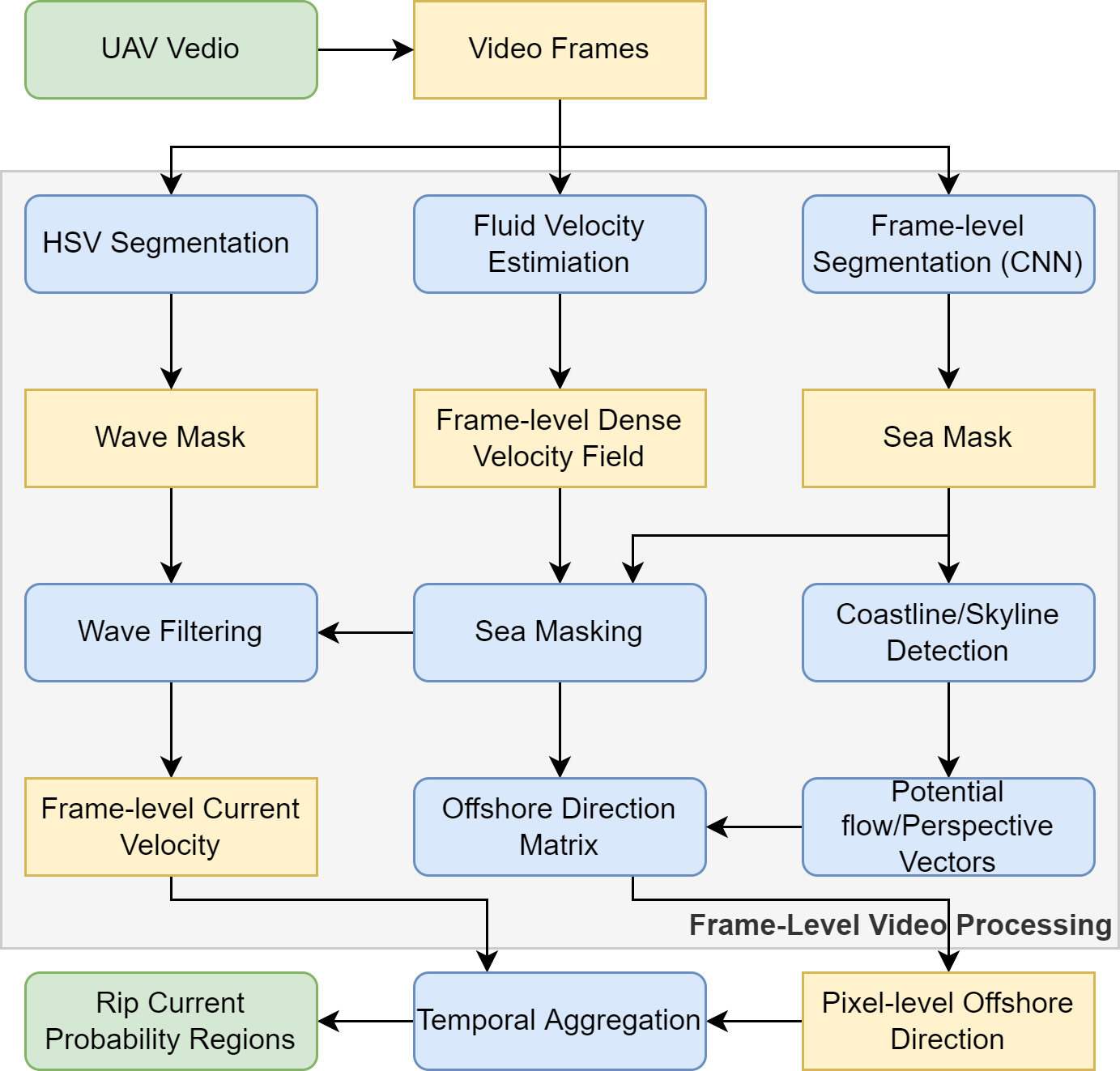}
	\caption{The proposed framework for UAV-video-based rip current detection}
	\label{fig:framework}
\end{figure}

Figure \ref{fig:framework} illustrates the rip current detection framework proposed in this paper based on UAV video analysis. The framework first applies the optical flow estimation method to analyze the UAV video at the frame level and generate a 2-dimensional dense fluid velocity field from a series of adjacent frames. The convolution neural network and heuristic algorithms are then used to segment the sea water and waves in the frame. Based on these results, the coastline and skyline in the frame can be detected to calculate the offshore direction and fine-grained current velocity for each pixel. These values are aggregated using a novel temporal data fusion method for accurate rip current detection. The proposed framework can produce a likelihood of the existence of rip currents for each pixel in the video.

\subsection{Fluid Velocity Estimation Methods}
In the proposed framework, one of the crucial components for detecting rip currents from UAV videos is to estimate the velocity of seawater. Optical flow estimation methods~\cite{lucas-k, horn1981determining}, a widely applied technique for video-based velocity estimation, are adopted to produce the two-dimensional velocity field at the fine-grained level. Various optical flow estimation methods are available, and five variants are adopted in this study, which will be discussed in Section~\ref{sec:experiment}. Empirical studies are conducted to evaluate the suitability of these methods for rip current detection. The study revealed that inappropriate constrains of the optimization can adversely affect the detection performance.

\subsubsection{Optical Flow Estimation Problem Formulation}
A two-dimensional velocity field $V = (u, v)$ can be used to describe the optical flow, which represents the pixel-level motions of adjacent frames in a video. The optical flow is determined by minimizing the displaced frame difference (DFD) between a given frame and its leading frame, where $D$ is a metric for measuring the difference between the two frames.
\begin{equation}
D = I_{x}u+I_{y}v+I_{t},
\end{equation}
where $I_{x}$, $I_{y}$, and $I_{t}$ are the matrices of derivatives of the image intensity values along the $x$, $y$, and time dimensions respectively. In the context of fluid velocity estimation, the estimated velocity field can be regarded as an approximation of the fluid velocity.
Equation (\ref{eq:optical_flow}) defines a general formula for such an estimation.
\begin{equation}
	\arg\min_{V\in E}{F_d(V) + \gamma F_r(V)},
	\label{eq:optical_flow}
\end{equation}
where $E$ denotes the domain of the two-dimensional velocity field, $F_d(V)$ represents the motion modeling term that minimizes DFD, $F_r(V)$ is the regularization term for the optimization problem, and $\gamma>0$ is the Lagrange multiplier used to balance the motion modeling and regularization terms. Various methods have been proposed to solve this optimization problem and obtain the optical flow estimation.

\subsubsection{Lucas-Kanade Optical Flow Estimation}
One classic method for solving the optical flow estimation problem is the Lucas-Kanade method~\cite{lucas-k}. This method assumes that the shift of image content is small and approximately constant in a local region between two adjacent frames. Thus, the motion modeling term $F_d(V_t)$ can be defined as:
\begin{equation}
	F_d(V_t) = \sum_{i=1}^{W}\sum_{j=1}^H{D_{i,j}^2},
\end{equation}
where $W$ and $H$ are the width and height of the frame, $D_{i,j}$ is the DFD value at pixel $(i,j)$, and no regularization term is applied for Lucas-Kanade method. This assumption is generally correct since the optical flow in successive frames tends to have similar direction and energy.

The Lucas-Kanade method solves the least squares principle and provides the following solution:
\begin{equation}
	V = (A^TA)^{-1}A^Tb,
\end{equation}
where $A=[I^*_x, I^*_y]$, $b = -I^*_t$, and $I^*_{x}$, $I^*_{y}$, and $I^*_{t}$ are the vectorized derivatives of the image intensity values along the $x$, $y$, and time dimensions, respectively.

\subsubsection{Horn-Schunck Optical Flow Estimation}
Incorporating global constraints for optical flow estimation can improve the performance of the method in scenarios where flow over the entire frame is smooth, as proposed by Horn and Schunck~\cite{horn1981determining}. Thus, the optimization problem in Equation (\ref{eq:optical_flow}) can be modified as follows:
\begin{equation}
	\arg\min_{V\in E}\iint D^2+\gamma\left ( \left | \triangledown u \right |^{2}+\left | \triangledown v \right |^{2} \right )dxdy
\end{equation}
Unlike the Lucas-Kanade method, which focuses more on local constraints, the Horn-Schunck method takes a holistic and global approach to optical flow estimation. This method is particularly suitable for estimating fluid velocity, as physical constraints must be satisfied in the real world. For instance, detecting flash rip currents, which are not continuous on the ocean surface, can be challenging. The image contents in the local region might not be constant, and the Lucas-Kanade method's assumption may not always hold in such cases.

\subsubsection{Typhoon}

In recent years, there has been a growing interest in using wavelet-based methods for optical flow estimation to estimate fluid velocity~\cite{derian2015wavelet}. One such method is Typhoon, which solves the dense variational optical flow problem by estimating the entire vector field for each pixel of the input images simultaneously~\cite{typhoon}. However, compared to traditional motion fields, the DFD function of Typhoon is non-linear, which makes the minimization process complex. Although this non-linear property of Typhoon allows it to compute large displacements, it also makes it difficult to minimize.

To overcome this challenge, Typhoon incorporates wavelet bases to facilitate the non-linear minimization~\cite{typhoon}. By providing a multiscale representation of the motion field, Typhoon can iteratively estimate the movement between images from rough to high precision. In summary, Typhoon is a suitable method for capturing surface waves and currents, especially in the ocean. However, it may not accurately capture rip current optical flow that flashes among waves due to the wavelet-based implementation.

\subsubsection{High-Order Regularization (HOR) for Optical Flow Estimation}
Building upon the Typhoon algorithm, we can leverage wavelet bases to improve sensitivity to surface waves and currents. However, this high sensitivity combined with the denoise method can lead to filtering of the majority of the optical flow of rip currents. To address this issue, we incorporate wavelet-based high-order regularization \cite{kadri2013divergence} to smooth the currents and facilitate deviations caused by brightness and abnormal motions of interfering objects such as seagulls.

By combining wavelet-based HOR with primary optical flow estimation methods such as Lucas-Kanade and Horn-Schunck, our algorithm can effectively compute smooth optical flow estimation for most currents on the ocean surface. This estimation enables better rip current detection through our evaluation method.

\subsection{Sea Water and Wave Segmentation}
\begin{figure}
	\centering
	\subfloat[Original Frame]{\includegraphics[width=0.2\textwidth]{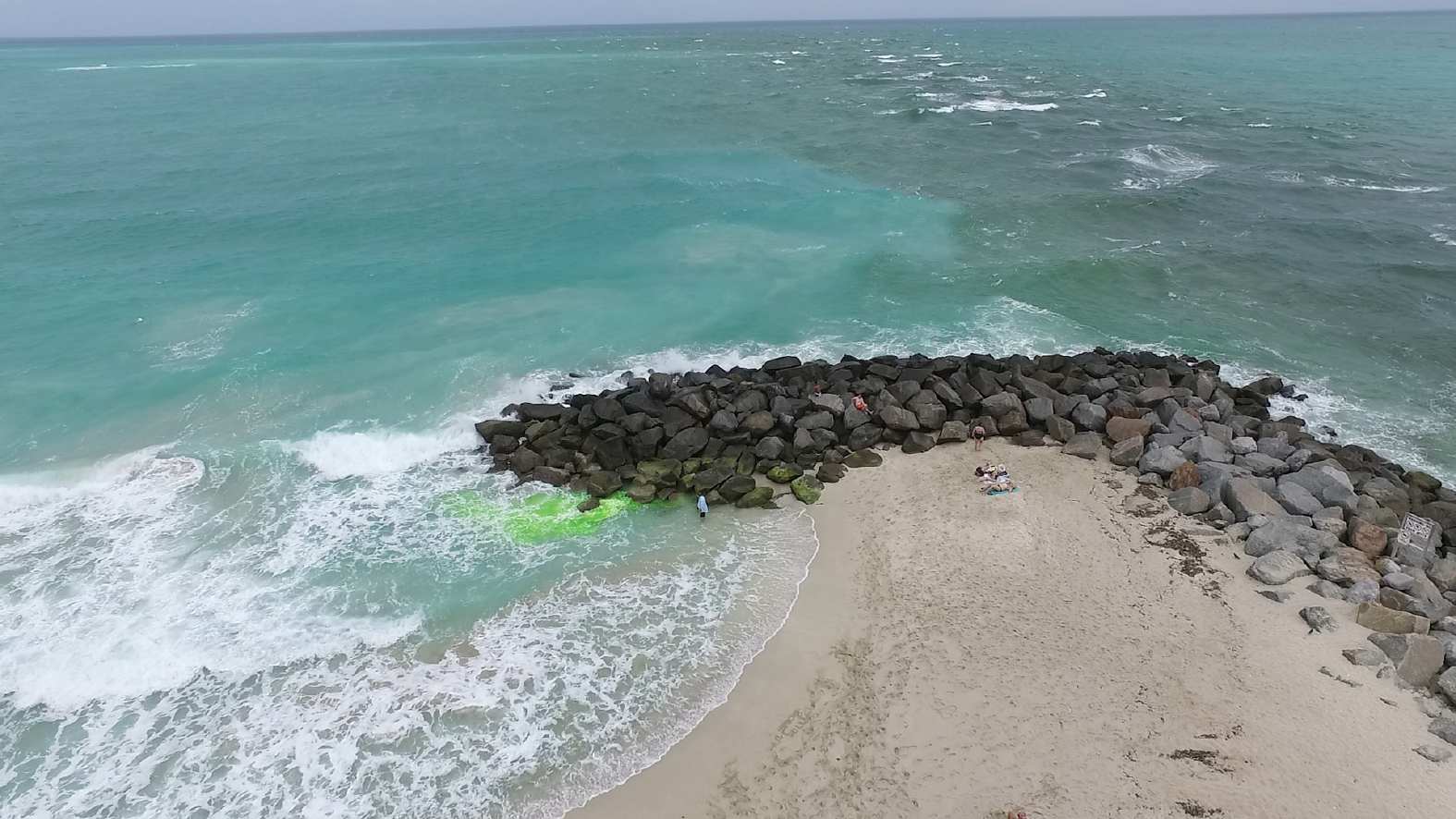}}\quad
	\subfloat[Shore Area Mask]{\includegraphics[width=0.2\textwidth]{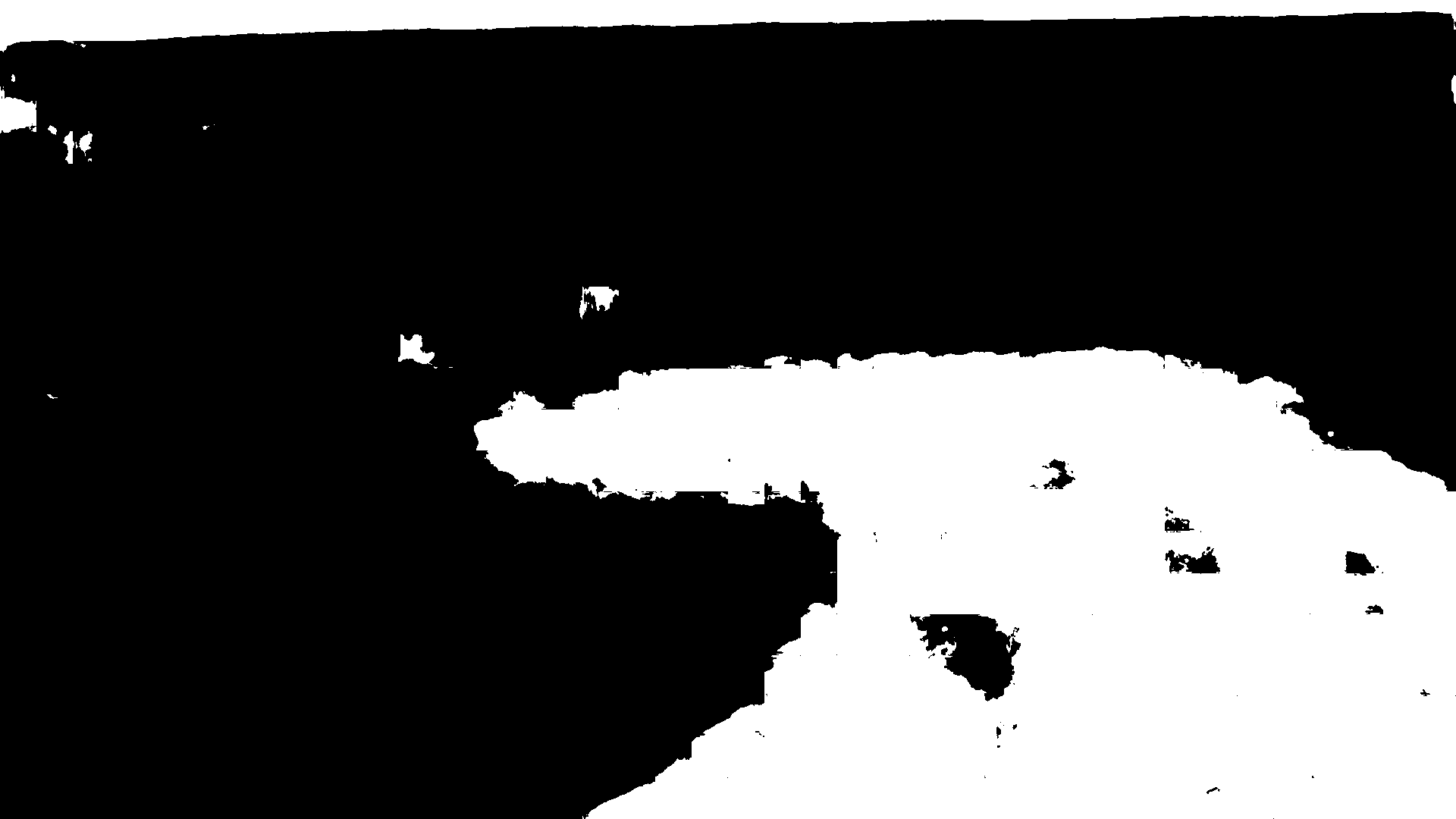}}\quad
	\subfloat[Wave Mask]{\includegraphics[width=0.2\textwidth]{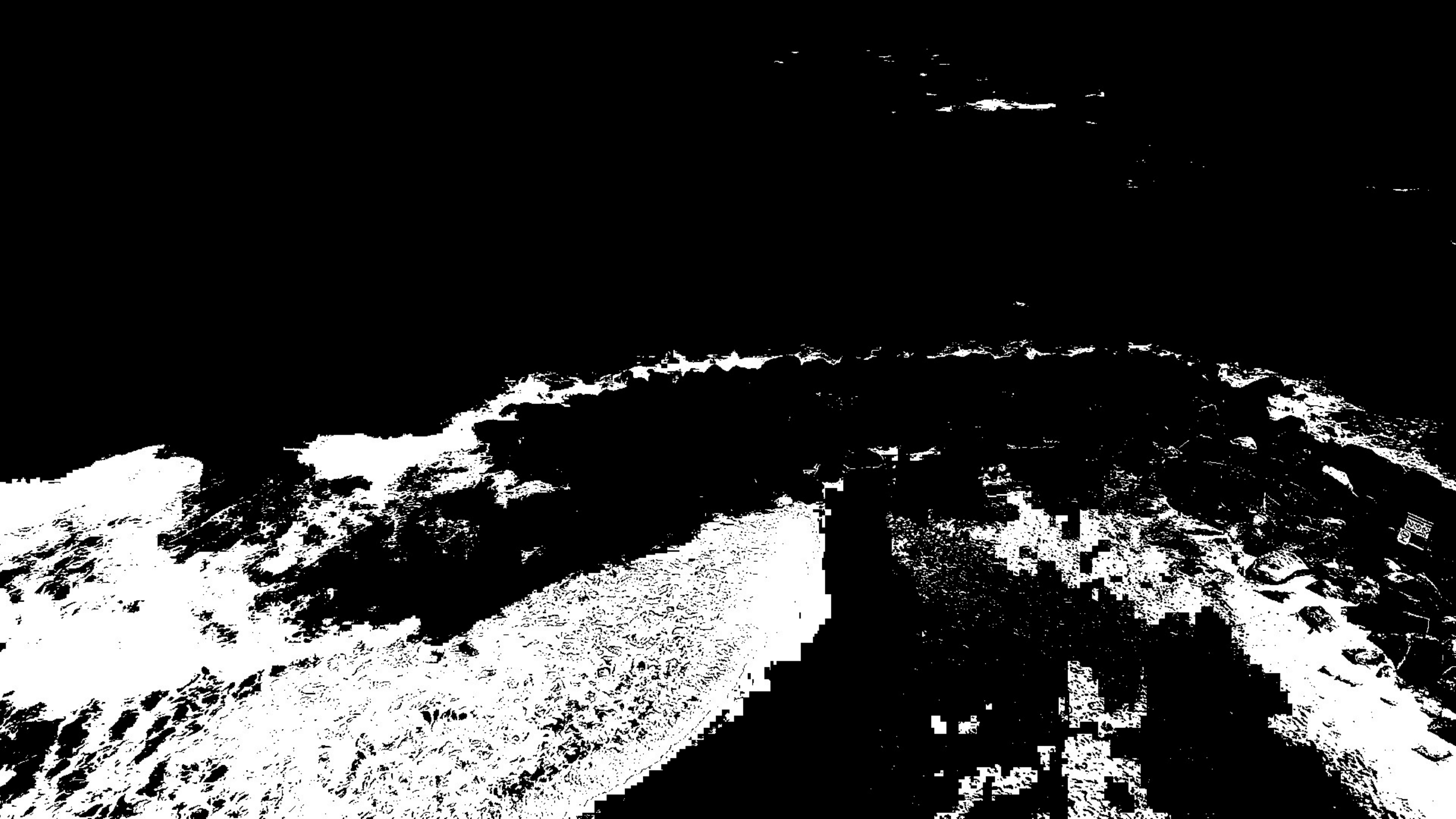}}\quad
	\subfloat[Combined Mask]{\includegraphics[width=0.2\textwidth]{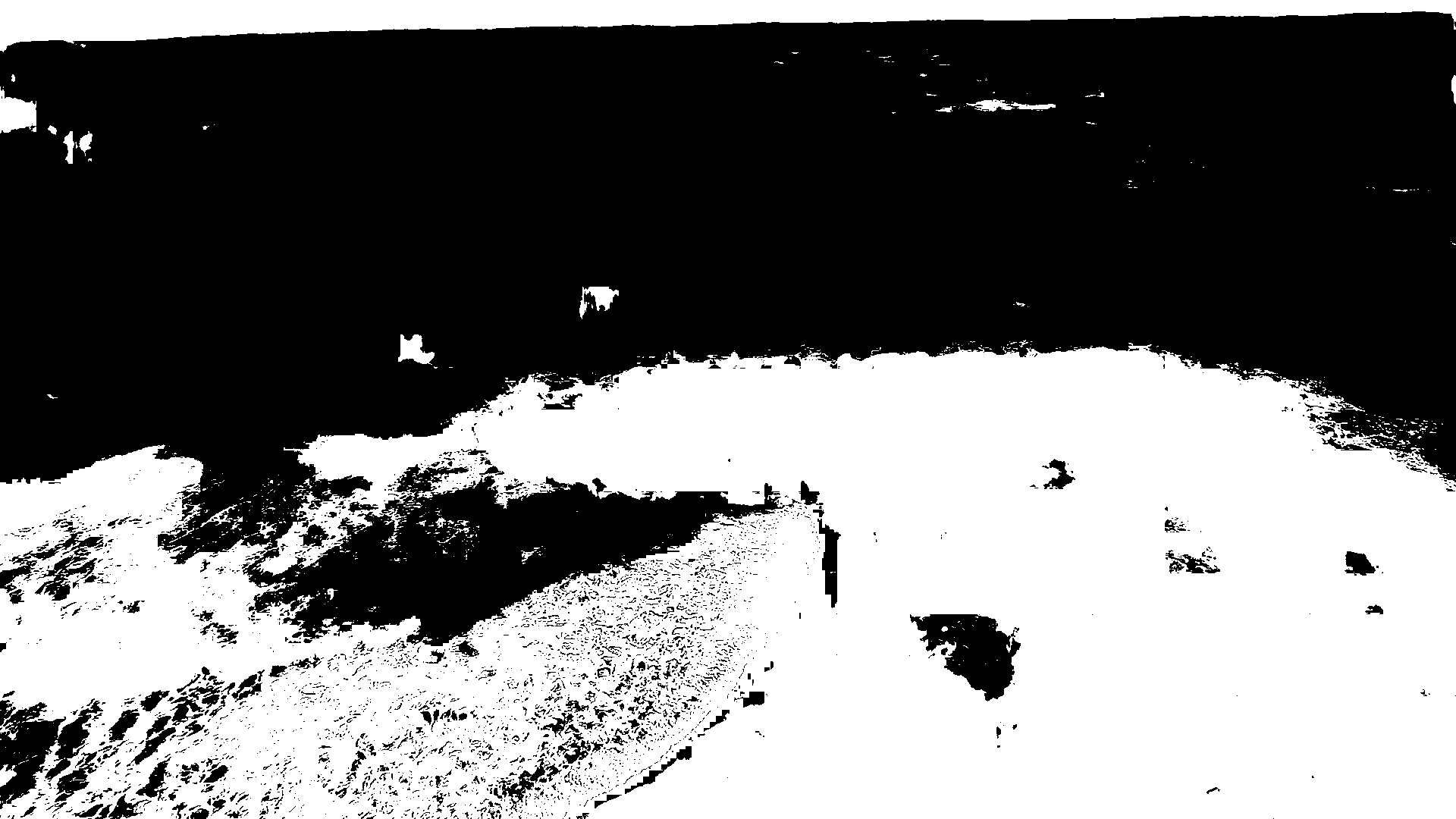}}
	\caption{Example of frame-level sea water and wave segmentation}
	\label{fig:seg}
\end{figure}
As shown in Fig.~\ref{fig:seg}(a), the UAV videos capture not only the sea surface but also the shore areas, making it necessary to segment out the shore areas to remove possible interference from these environments. However, due to air turbulence and wind, the UAV can be unstable and shaking while recording the videos, resulting in noise on the shore, sky, and some other regions. As a result, the first-order Lucas-Kanade algorithm can produce misleading results.

Moreover, in UAV videos, the motion of waves on the sea surface usually dominates the motions captured by optical flow estimation due to their better visual visibility and the presence of vortex in their tails, which shows similar directions as the offshore directions. To better capture subtle currents and detect rip currents, we propose applying wave segmentation in our framework to suppress the impact of vortex in the waves.

To accurately estimate rip currents from UAV videos, it is essential to segment out the shore areas and suppress the impact of waves in the optical flow estimation. To achieve this, we propose a framework that applies a Convolutional Neural Network (CNN) with a U-Net architecture~\cite{RonnebergerFB15} and ResNet-50 backbone~\cite{HeZRS16} for sea water segmentation. This CNN-based model is trained with a custom dataset of images taken by the UAV in near-shore areas. CNN has shown superior performance compared to other algorithms for image segmentation in many domains. The shore area mask is generated using the CNN-based model for each frame in the near-shore UAV videos, which covers most of the non-water areas and can be generalized to various scenes Fig.~\ref{fig:seg}(b).

To remove the noise in velocity fields generated by the optical flow estimation methods due to the waves, we apply wave segmentation. Since the color of the foam is similar to white or light red under sunlight reflection regardless of time and location, it can be easily distinguished from the other pixels of the sea water based on the saturation and color value. The wave mask Fig.\ref{fig:seg}(c) can be generated using the pre-determined ranges in the HSV color space. In the end, the combined mask Fig.\ref{fig:seg}(d) is generated as the union of the shore area and wave masks. Thus, using the combined masks and frame-level velocity fields, the offshore directions can be estimated, and the noise in velocity fields generated by the optical flow estimation methods can be removed.

\subsection{Off-Shore Direction Calculation}
Assuming a resolution of $n \times m$ for a frame containing the distance between each pixel point $(x,y)$ and the shoreline, the directional derivative can be calculated using the gradient of the distance matrix to determine the direction of motion for each pixel. By treating the calculated horizontal gradients $(u)$ as rows and vertical gradients $(v)$ as columns, a vector can be created for each pixel on the ocean surface. Using camera projection in 3-D space, the direction of each point on the ocean surface can be accurately determined in real-world coordinates by mapping from a spherical planet to 2-D frames. This is achieved through near-local isometric embedding techniques on the sphere, with the shoreline and skyline providing reference points to assess the range of directions offshore. Through this pipeline, it is possible to determine whether a pixel on the ocean surface has an off-shore direction. The proposed pipeline can be simplified and divided into the following steps during implementation.

\subsubsection{Almost Local-Isometric Embedding Modeling on Sphere}
Assume the screen is located at $(a,0,0)$ with size $n \times m$, the coordinate for 4 corners are $(a,\pm \frac{n}{2},\pm \frac{m}{2})$.

The distance between the camera and the ground is $h$ and the radius of the earth is $R$.

The equation of earth $E$ is
\begin{equation}
    \begin{aligned}
    E : (x-(h+R)\cos(\theta)\sin(\psi))^2 \\
    +(y-(h+R)\sin(\theta)\sin(\psi)) \\
    +(z-(h+R)\cos(\psi))^2=R^2
\end{aligned}
\end{equation}
and assume $\cos(\theta)\sin(\psi)>0$.

Pick a point $p$ on the screen $p=(a,u,v)$ with $-\frac{n}{2} \leq u \leq \frac{n}{2},-\frac{m}{2}\leq v \leq \frac{m}{2}$, we could find a line passing through it:
\begin{equation}
    l(t)=(at,ut,vt)
\end{equation}

We try to find the intersection of the line with the earth:

\begin{equation}
\begin{aligned}
    &(at-(h+R)\cos(\theta)\sin(\psi))^2+ \\
    &(ut-(h+R)\sin(\theta)\sin(\psi))+ \\
    &(vt-(h+R)\cos(\psi))^2=R^2
\end{aligned}
\end{equation}

\begin{equation}
\begin{aligned}
&(a^2+u^2+v^2)t^2-2(h+R)(a\cos(\theta)\sin(\psi)+\\
    &u\sin(\theta)\sin(\psi)+v\cos(\psi))t+h(h+2R)=0
\end{aligned}
\end{equation}

The discriminant of this quadratic equation is
\begin{equation}
\begin{aligned}
    \Delta=&4(h+R)^2(a\cos(\theta)\sin(\psi)+u\sin(\theta)\sin(\psi)+\\
    &v\cos(\psi))^2-4(a^2+u^2+v^2)h(h+2R)
\end{aligned}
\end{equation}

If we let $\Delta=0$, it gives the equation of the horizontal line $(u,v)$ with conditions $-\frac{n}{2} \leq u \leq \frac{n}{2}, -\frac{m}{2} \leq v \leq \frac{m}{2}$.

Assume the line intersects with the earth, then we have

\begin{equation}
\begin{aligned}
    t = &\frac{2(h+R)\left(a\cos(\theta)\sin(\psi) + u\sin(\theta)\sin(\psi) + v\cos(\psi)\right)}{2(a^2+u^2+v^2)} \\
    &\pm \frac{\sqrt{\Delta}}{2(a^2+u^2+v^2)}
\end{aligned}
\end{equation}

Obviously, we should require $\Delta \geq 0$ and choose
\begin{equation}
\begin{aligned}
    t(u,v) = &\frac{2(h+R)}{2(a^2+u^2+v^2)}\left(a\cos(\theta)\sin(\psi) + u\sin(\theta)\sin(\psi) \right. \\
    &\left. + v\cos(\psi)\right) - \frac{\sqrt{\Delta}}{2(a^2+u^2+v^2)}
\end{aligned}
\end{equation}

We have a map $f$ from domain 
\begin{equation}
\begin{aligned}
D=\{(u,v):u \in [-\frac{n}{2},\frac{n}{2}],v \in [-\frac{m}{2},\frac{m}{2}], \Delta \geq 0\}
\end{aligned}
\end{equation}
to $E$ given by
\begin{equation}
\begin{aligned}
    f : (u,v) \to (dt(u,v),ut(u,v),vt(u,v))=p(u,v)
\end{aligned}
\end{equation}
where $t(u,v)$ is given as above.

Since there is no local isometry between the sphere and plane. We try to find a map from $f(D)$ to $\mathbb{R}^2$ that minimizes the change of distance(?).

Given two points $q_1=(x_1,y_1,z_1)$, $q_2=(x_2,y_2,z_2)$ on sphere $S:x^2+y^2+z^2=R^2$, the great \begin{equation}
\begin{aligned}
    d_c(q_1,q_2)=R \cdot \min\{&arccos(\frac{q_1 \cdot q_2}{R^2}), \\
    &arccos(\frac{- q_1 \cdot q_2}{R^2})\}
\end{aligned}
\end{equation}

We could write the Euclidean coordinates into spherical coordinates with the chosen north pole and 0 longitude angle:
\begin{equation}
\begin{aligned}
    &x=R \cos(\phi)\sin(\delta)\\
    &y=R \sin(\phi)\sin(\delta)\\
    &z=R \cos(\delta)
\end{aligned}
\end{equation}

and write down the map $g_q$ from the $S_{z \geq 0}$ to plane $\mathbb{R}^2$ where $q$ is the north pole.

\begin{equation}
\begin{aligned}
    g_q(\phi,\delta)=(R\cdot \delta \cdot \cos(\phi),R \cdot \delta  \cdot \sin(\phi))
\end{aligned}
\end{equation}

Obviously, the map $g$ does not depend on the choice of 0 longitude angle and depends only on the choice of the north pole $q$.

We may translate the same metric to $E$ and define function
\begin{equation}
\begin{aligned}
    W(q)=\iint_{f(D)} |d_c(q_1,q_2)-d(g_q(q_1),g_q(q_2))|dq_1dq_2
\end{aligned}
\end{equation}
where $q \in f(D)$.
Since $f(D)^2$ is bounded and closed, function $W$ is continuous, thus we have an absolute minimum on $f(D)^2$ and assume we find a minimizer $q_0 \in f(D)$.

With initial data $a,h,R,\theta,\psi$ and the minimizer $q_0$, we have a map from $h=g_{q_0}\circ f: D \to \mathbb{R}^2$, that is "almost local-isometric".

\subsubsection{The Estimation of Local Offshore Direction}
\begin{figure}
    \centering
    \includegraphics[width=0.25\textwidth]{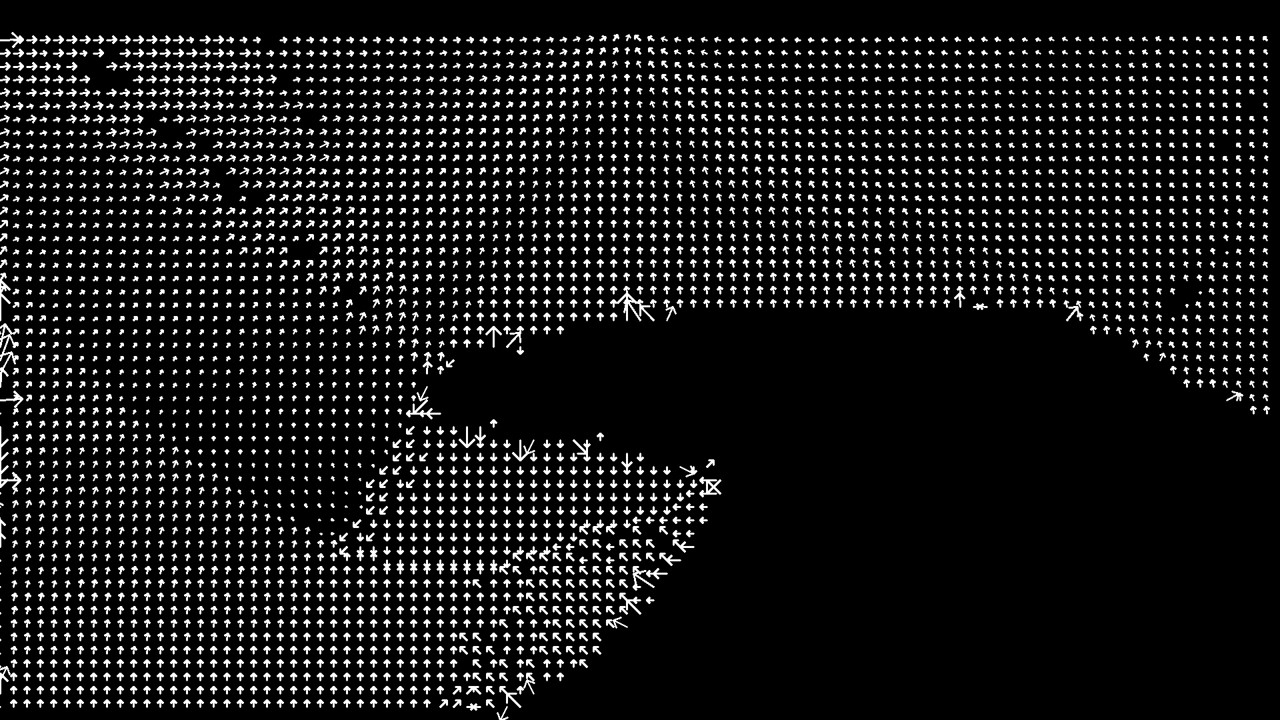}
    \caption{The aggregated pixel-level offshore direction}
    \label{fig:offshore}
\end{figure}

After obtaining an almost local-isometric projection of videos onto the actual Earth's ocean surface, we introduce a pipeline designed to identify shoreline information within the projected almost local-isometric frame masks. Gaussian filtering is used to reduce noise in the mask, followed by the application of the Canny edge detector to obtain the edge image from the mask. Subsequently, the coastline is determined using the detected edge image. If the coastline is discontinuous or outside the edge image boundary, the bottom edge of the image is assumed to be part of the coastline for subsequent calculations. Once the shoreline information is detected, we can formulate a matrix $S$ that represents the distances between each pixel in the frame and the detected shoreline points ${(N_1, M_1), (N_2, M_2), ..., (N_n, M_n)}$ as follows:
\begin{equation}
	\scriptsize
	\begin{aligned}
		S_{i,j}=\min \left\{\sqrt{(N_{1}-i)^2+(M_{1}-j)^2}, \sqrt{(N_{2}-i)^2+(M_{2}-j)^2}, \right.\\
		\phantom{=\;\;}
		\left...., \sqrt{(N_{n}-i)^2+(M_{n}-j)^2} \right\}
	\end{aligned}
\end{equation}
Using the distance matrix $S$, we can compute the potential flow in the frame. The results are then masked with the generated shore area mask to obtain the vectors of pixel points on the sea surface in the potential flow, which are referred to as offshore directions $OV_{i,j}$. The offshore direction matrix based on the segmentation masks can be visualized as a flow map, as shown in Fig.~\ref{fig:offshore}



\subsection{Rip Current Detection}
Even when the estimated current velocity and offshore direction are obtained, it is hard to evaluate whether there is a rip current using a single frame. Therefore,  temporal data fusion is introduced to utilize the estimated velocity fields over time, even after obtaining the estimated current velocity and offshore direction. For the $t$-th frame $I_t$ in the video, let $V_{t,i,j} = (u_{t,i,j}, v_{t,i,j})$ be the estimated velocity field using any optical flow estimation method, $M^g_t$ be the ground mask, and $M^c_t$ be the combined mask. Given the offshore direction $OV_{t,i,j}$, the rip currents can be defined as any current directed away from the seashore, with the angle between the current direction and $OV_{t,i,j}$ within the range of $(-60^{\circ}, +60^{\circ})$. 

Since have an almost local-isometric map from $h=g_{q_0}\circ f: D \to \mathbb{R}^2$, and this map $h$ is differentiable and injective. Given a vector field $V$ on domain $D$, it will induce a vector field $dh(V)$ on $h(D)$ where
\begin{equation}
\begin{aligned}
    (dh(V))(h(p))=dh_{p}(V(p))=(h\circ\gamma)'(0)
\end{aligned}
\end{equation}
where $\gamma$ is a curve on $D$ such that $\gamma(0)=p$ and $\gamma'(0)=V(p)$.

Given two vector fields $OV_{t,i,j}$ which is offshore direction, and $V_{t,i,j}$ which is pixel optical flow measurement on $D$, the angle between $dh(OV_{t,i,j})$ and $dh(V_{t,i,j})$ at point $h(p)$ is given by
\begin{equation}
\begin{aligned}
    \Theta(h(p))_{t,i,j}=arccos(\frac{dh_{p}(OV_{t,i,j}(p)) \cdot dh_p(V_{t,i,j}(p))}{|dh_{p}(OV_{t,i,j}(p))|\cdot |dh_p(V_{t,i,j}(p))|})
\end{aligned}
\end{equation}
where $p \in D$. Then we can compare the $\Theta(h(p))_{t,i,j}$ with the range of $(-60^{\circ}, +60^{\circ})$ to detect whether this pixel in this frame is offshore or not.

Since various factors could cause fluid velocity towards the offshore direction, it is unreliable to estimate the rip current regions based on the results of a single frame. Thus, to obtain a fine-grained estimation of the rip current regions, it is essential to aggregate the results from the frames of the video and fuse the data by utilizing the temporal information. To accomplish this, the rip current likelihood matrix $L$ is proposed, defined as the frequency of each pixel being within the rip current regions for a given time period $T$, using the recursive formula below:
\begin{equation}
	\left\{
	\begin{aligned}
		&L_{t,i,j} = L_{t-1,i,j} + 1  &&\text{if}\Theta(h(p))_{t,i,j} \in (-60^{\circ}, +60^{\circ})\\ 
		&L_{t,i,j} = L_{t-1,i,j} &&\text{otherwise}
	\end{aligned}
	\right.,
\end{equation}
where $L_{t,i,j}$ is the likelihood matrix of pixel $(i,j)$ for the period of $[0, t]$, and the final likelihood matrix is defined as $L_{T} = {L_{T, i, j}}{i\in[1, W], j\in[1, H]}$. Specifically, $L{0, i, j}=0, \forall i\in[1, W], j\in[1, H]$. The higher the value of $L_{T,i,j}$ for the pixel $(i,j)$, the higher the probability that there is a rip current at the location of this pixel.

\begin{table*}
	\begin{center}
		\caption{The Summary of Near-Shore UAV Video MetaData}
		\label{tab1}
		\begin{tabular}{|c|c|c|c|c|c|}
			\hline
			\textbf{Video Name}	& \textbf{Flight Altitude} &\textbf{Location} &\textbf{Start Time} &\textbf{Characteristics of the Area}&\textbf{Rip Current}\\
			\hline
			Palm Beach	&50m &26.793436, -80.031638 &3:34PM April 27, 2016 &Clear Water, Rock Outcropping&Captured\\ \hline
			Haulover	&70m	&25.900867, -80.121667 &2:09PM April 6th, 2016 &Complex Rocky Jetty&Captured \\ \hline
			Ocean Reef Park &Varying &26.792321, -80.032065 &9:07AM June 18th, 2016 & Clear Water, Rock Outcropping&Not Captured \\ \hline
			South Beach 1&Varying &26.792321, -80.032065 &1:11PM Sept. 29th, 2018 & Sargassum Seaweed Present&Not Captured \\ \hline
			South Beach 2&45m &25.791486, -80.127193 &11:51AM Feb. 21st, 2021 & Bar Gap Present &Captured \\ \hline
			YouTube Videos  &NA &NA &NA &NA & Captured \\ \hline
		\end{tabular}
	\end{center}
	\vspace{-1em}
\end{table*}

\section{Experiment Results}\label{sec:experiment}

\subsection{Near-Shore UAV Video Data Acquisition}
We present a dataset of four UAV-recorded videos of the near-shore area in the United States, two of which capture rip currents. All videos were captured at 60Hz frequency using a DJI Phantom 3 Advanced Quadcopter UAV, equipped with a Full HD $1920 \times 1080p$ camera and a 3-axis stabilization gimbal to reduce image blurring. Prior to processing by the optical flow estimation methods, all video frames were converted to gray-scale. Table~\ref{tab1} provides metadata for the videos.

One of the two videos that captured a rip current was recorded at Riviera Beach, located in Palm Beach County, Florida. This beach features a structurally-controlled rip current that forms through an outcropping of coquina limestone rock called the Anastasia formation. At low tides, the rock outcropping is almost fully exposed, and fluorescein dye delineates the rip current as it flows seaward through the rocks. The flow characteristics of this rip current showed a relatively straight-flowing rip that terminates at the outer edge of the surf zone. The video was captured on April 27th, 2016, during good weather conditions, with significant wave height of 0.6 m from the east-southeast and wind speeds of only 10 kts from the east~\cite{leatherman2017rip, leatherman2017techniques}.

The other rip current video was recorded at Haulover Park, which is the largest surf beach park in Miami-Dade County and is situated north of Miami Beach. The park features a jetty structure with a beach-parallel spur (breakwater) at the end that acts as a barrier against waves. Rip currents form in response to longshore currents generated by obliquely-approaching waves from the northeast. The southward-flowing longshore current is deflected by the jetty to form a structurally-controlled rip that flows to the north and then offshore. The video was recorded on April 6th, 2016, in good weather conditions, with significant wave height of 0.8 m, spilling breakers, and a 6 second period, and wind speeds of 15 kts from the northeast~\cite{leatherman2017rip, leatherman2017techniques}.

The other two videos in our dataset were recorded at Ocean Reef Park and South Beach, both in Florida, and were used solely for training the proposed sea segmentation CNN model. To increase the diversity of video frames in the dataset, the UAV's location, pose, and flight altitude varied throughout the recordings. All four videos showcase different types of near-shore scenes, where the color of the sea water, beach, and rock, the objects in the near-shore areas, and the shape of the shores differ. Our proposed framework is limited to videos with clear vision of the sea surface and near-shore area, recorded under good weather conditions. However, in the future, we plan to extend our work to include videos captured under different weather conditions, such as rainy and stormy days.

\subsection{Fluid Velocity Estimation Performance Evaluation}
\subsubsection{Expert Rating Based on Quiver Plot}
\begin{figure}
	\centering
	\includegraphics[width=0.25\textwidth,trim={7cm 1cm 1cm 3cm},clip]{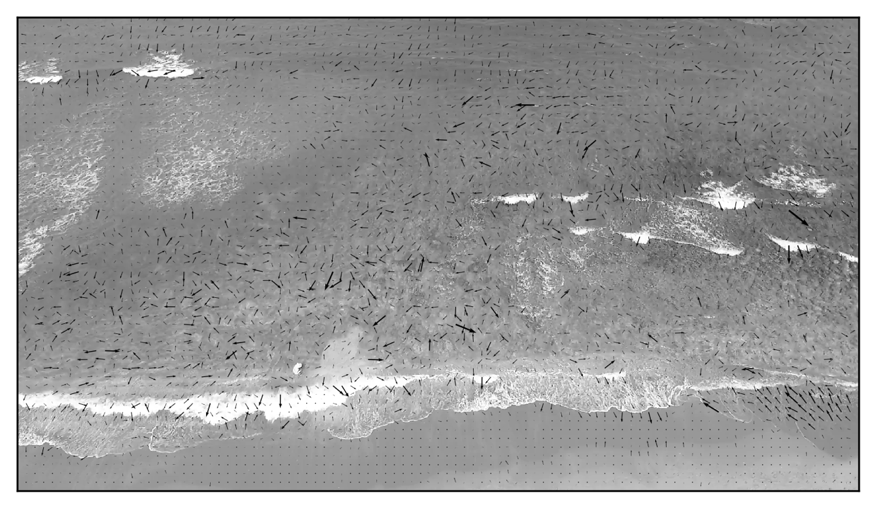}
	\caption{Example quiver plot for the Palm Beach video}
	\label{fig:quiver}
	\vspace{-1em}
\end{figure}
The quiver plot is a useful technique for visualizing two-dimensional velocity fields and aiding oceanographers in observing estimated fluid velocities. To generate a quiver plot based on the estimated velocity $V$ of a frame, the frame is initially split into blocks of size $N_W \times N_H$, where the size of the frame $W \times H$ is divided into $\frac{W}{N_W} \times \frac{H}{N_H}$ blocks. For each block, an average velocity is computed using the equation:
\begin{equation}
	\bar{V}_{block} = \frac{WH}{N_WN_H}\sum_{i=1}^{\frac{W}{N_W}}\sum_{j=1}^{\frac{H}{N_H}}{V_{i_0+i,j_0+j}},
\end{equation}
where $V_{i,j}$ is the velocity at pixel $(i,j)$, $(i_0, j_0)$ represents the index of the pixel in the left-upper corner of the block, and it is assumed that $\frac{W}{N_W}$ and $\frac{H}{N_H}$ are integers.

Once the average velocity for each block is computed, an arrow is superimposed on the frame image for each block, where the length of the arrow represents the speed, and the direction of the arrow indicates the direction of the velocity. An example of a quiver plot is shown in Fig.~\ref{fig:quiver}.

To visualize the velocity estimation and evaluate whether the estimated fluid velocity aligns with the actual rip current direction as measured in the field, quiver plots for all adjacent frames in the video are generated and concatenated to produce a quiver plot video. In the case of the Haulover and Palm Beach videos in the experiment, the Typhoon optical flow estimation method's quiver plot results are favored because it provides a cleaner velocity estimation compared to other algorithms. However, it should be noted that the Typhoon method tends to ignore the velocity of subtle currents and provides only an estimate for the waves with foam, which are visually identifiable.

\subsubsection{Virtual Drifter}

\begin{figure}
	\centering
	\subfloat[Initial positions]{
		\includegraphics[width=0.2\textwidth,trim={15cm 0 15cm 12cm},clip]{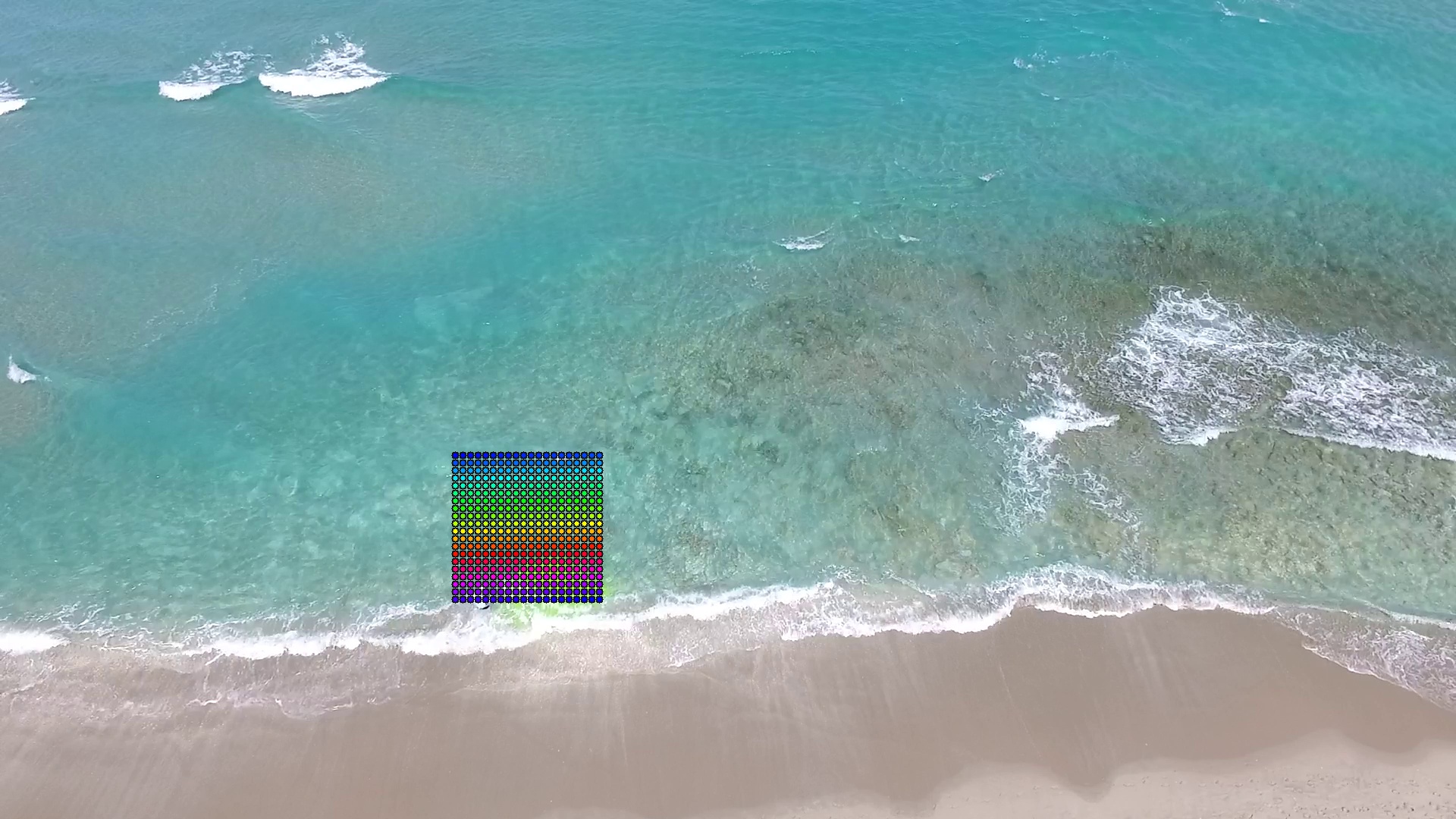}
	}
	\quad
	\subfloat[After 20-second simulation]{
		\includegraphics[width=0.2\textwidth,trim={15cm 0 15cm 12cm},clip]{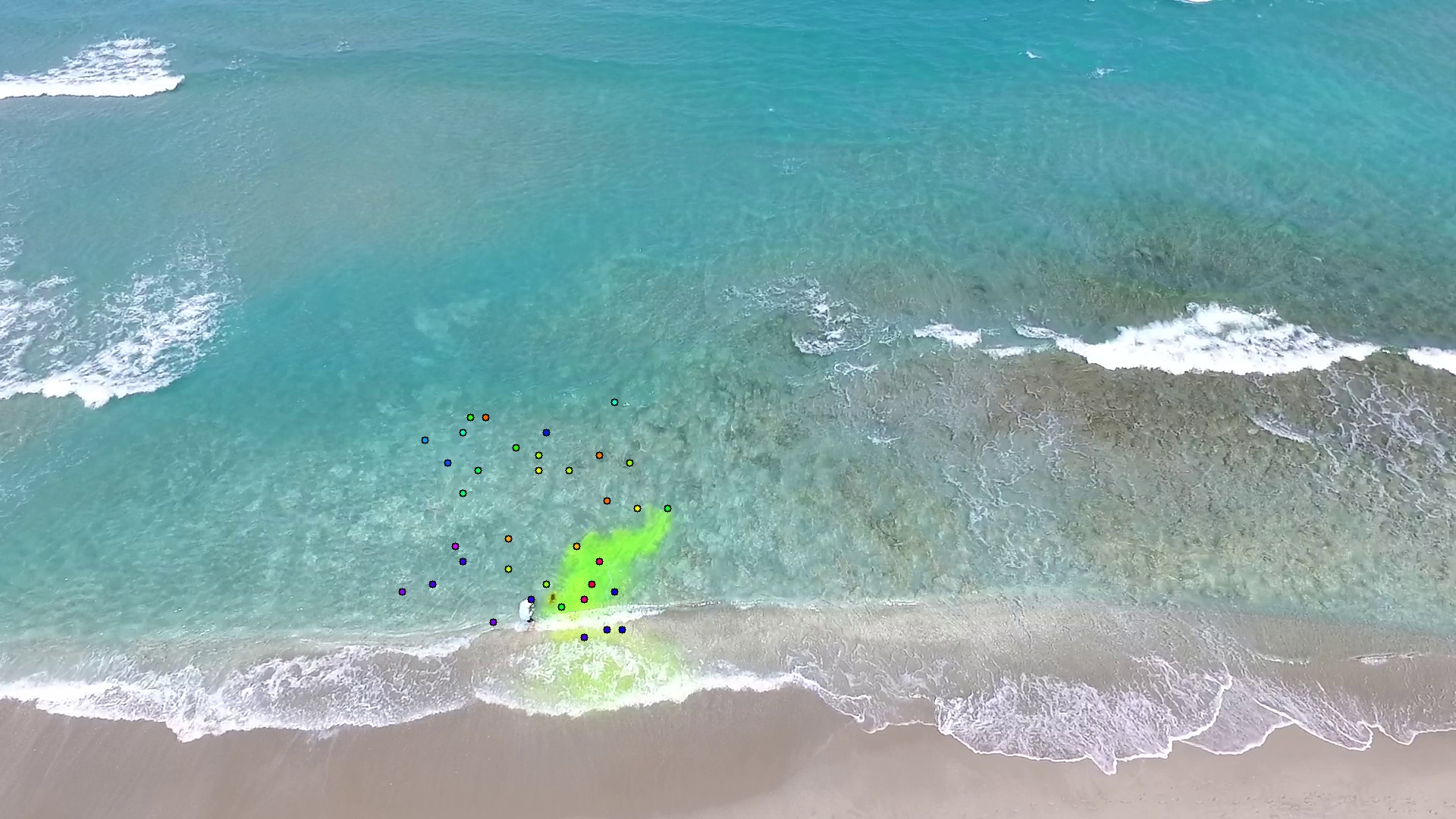}
	}
	\quad
	\subfloat[After 40-second simulation]{
		\includegraphics[width=0.2\textwidth,trim={15cm 0 15cm 12cm},clip]{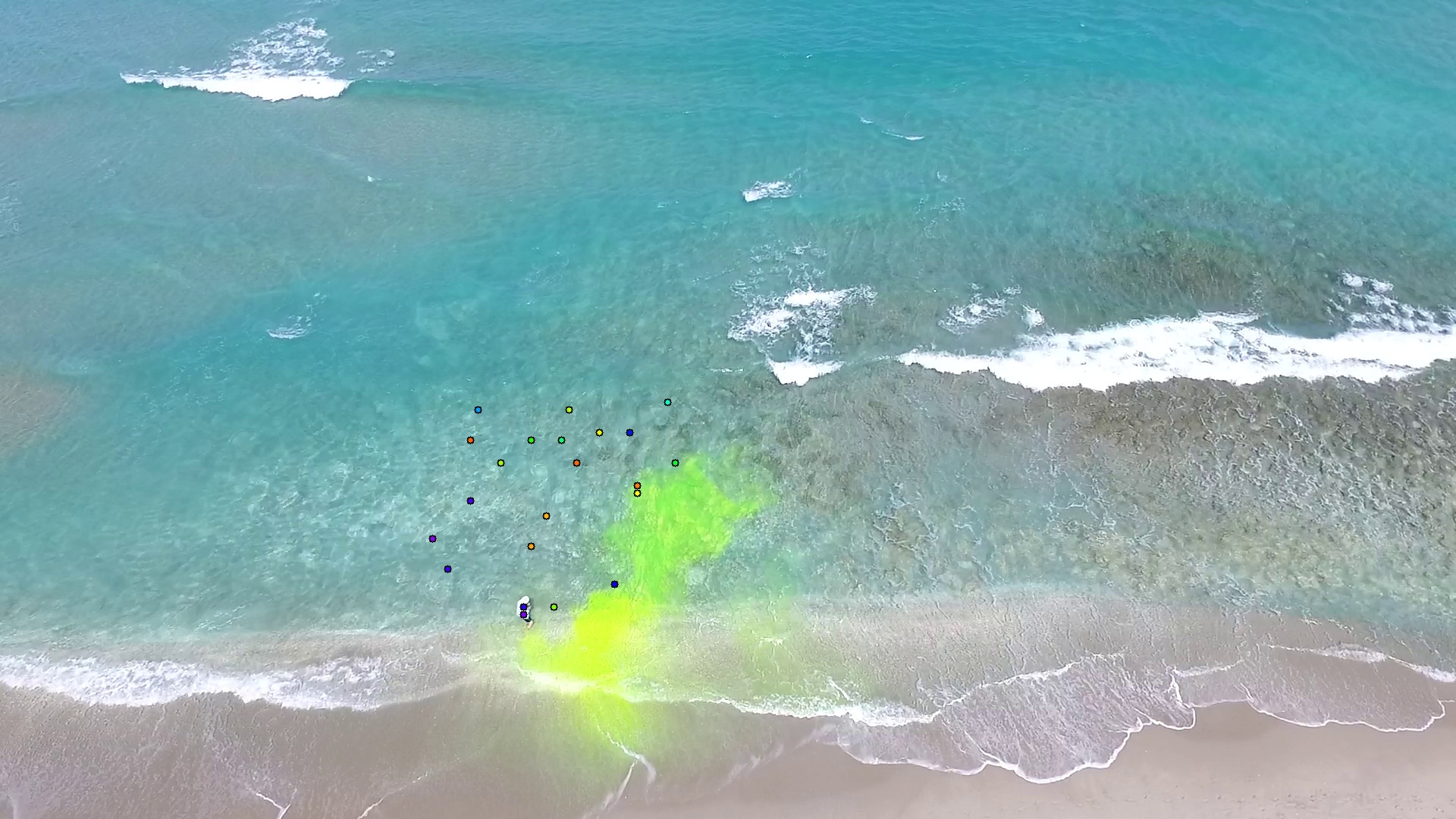}
}
	\quad
	\subfloat[After 60-second simulation]{
		\includegraphics[width=0.2\textwidth,trim={15cm 0 15cm 12cm},clip]{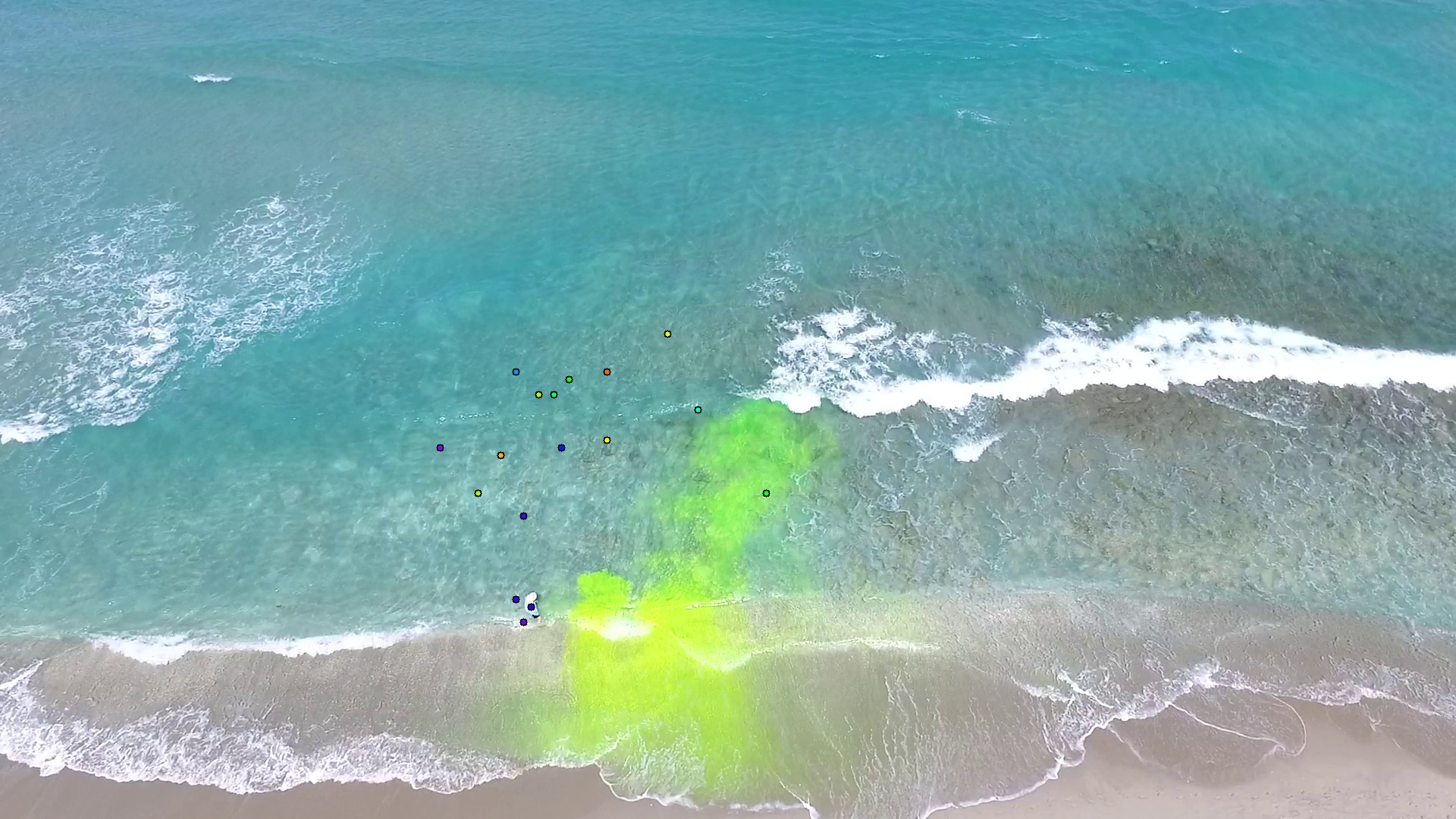}
}
	\caption{Positions of virtual drifters for the Palm Beach video}
	\label{fig:dye}
	\vspace{-1em}
\end{figure}

\begin{figure*}
	\centering
	\subfloat{
		\includegraphics[width=0.14\textwidth,trim={17.8cm 0 0 0},clip]{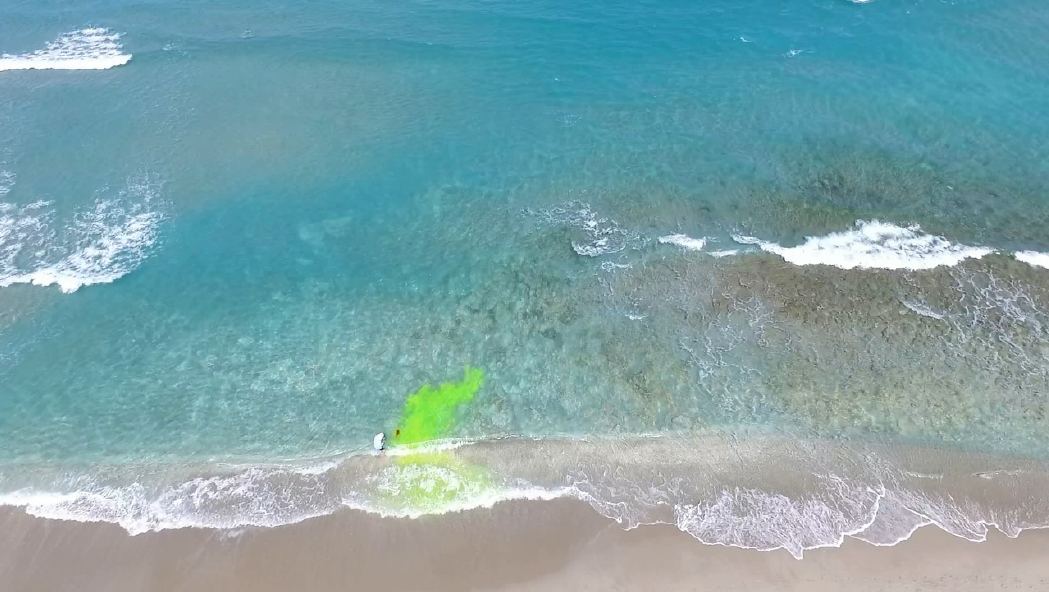}
	}
	\subfloat{
		\includegraphics[width=0.14\textwidth,trim={0 0 67.7cm 0},clip]{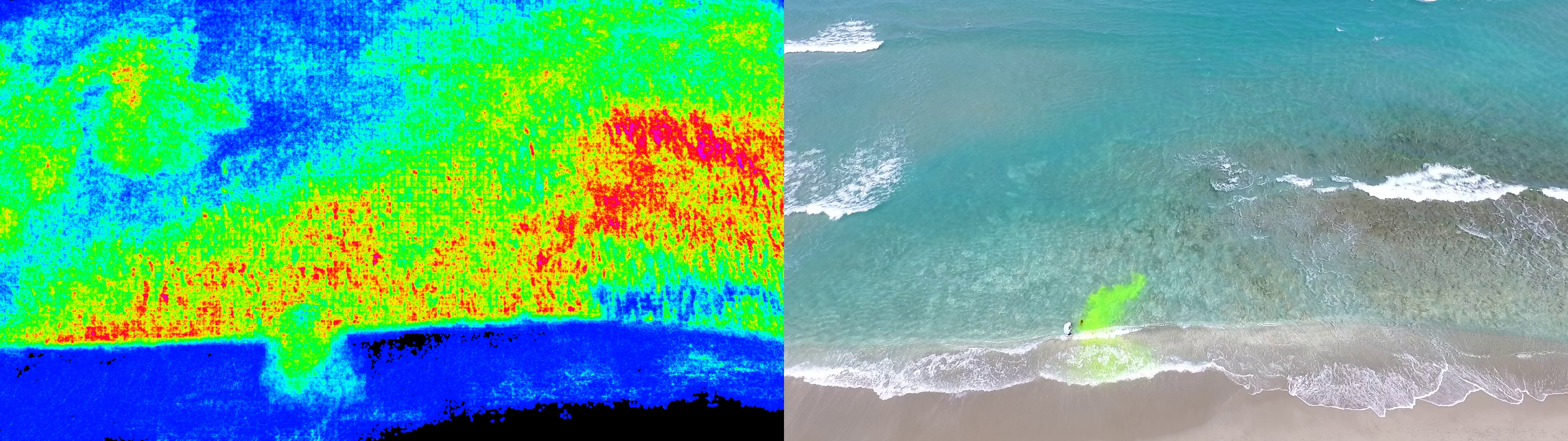}
	}
	\subfloat{
		\includegraphics[width=0.14\textwidth,trim={0 0 67.7cm 0},clip]{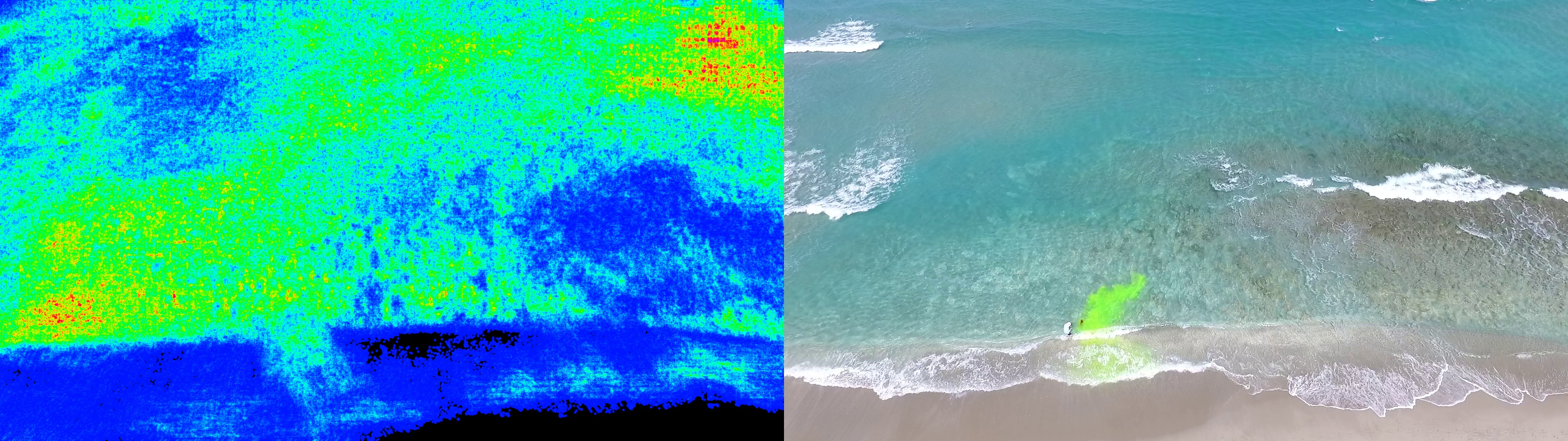}
	}
	\subfloat{
		\includegraphics[width=0.14\textwidth,trim={0 0 67.7cm 0},clip]{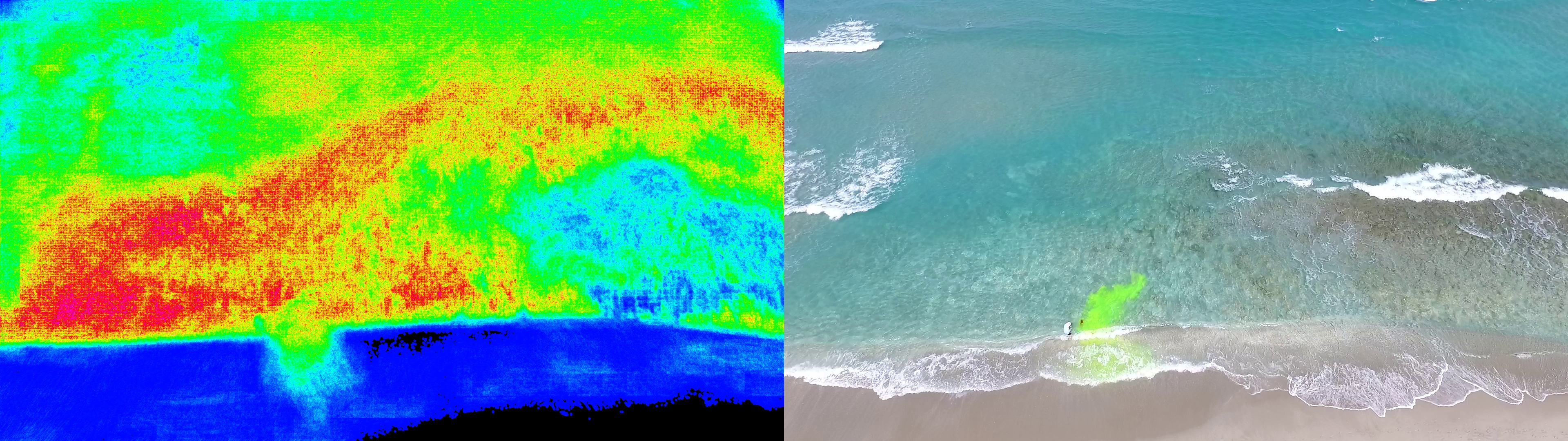}
	}
	\subfloat{
		\includegraphics[width=0.14\textwidth,trim={0 0 67.7cm 0},clip]{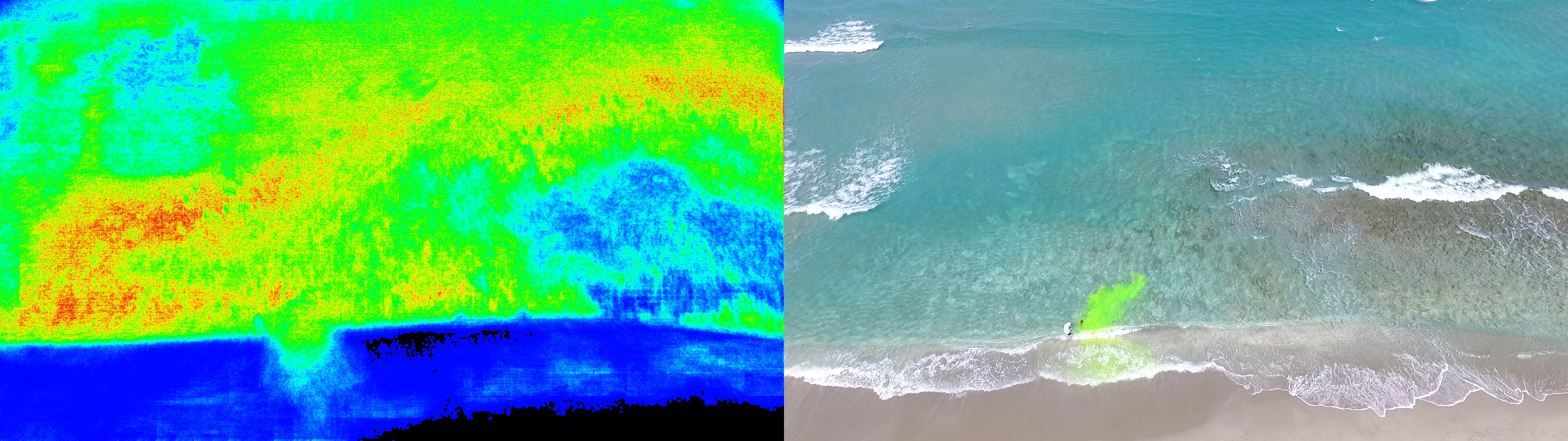}
	}
	\subfloat{
		\includegraphics[width=0.14\textwidth,trim={0 0 67.7cm 0},clip]{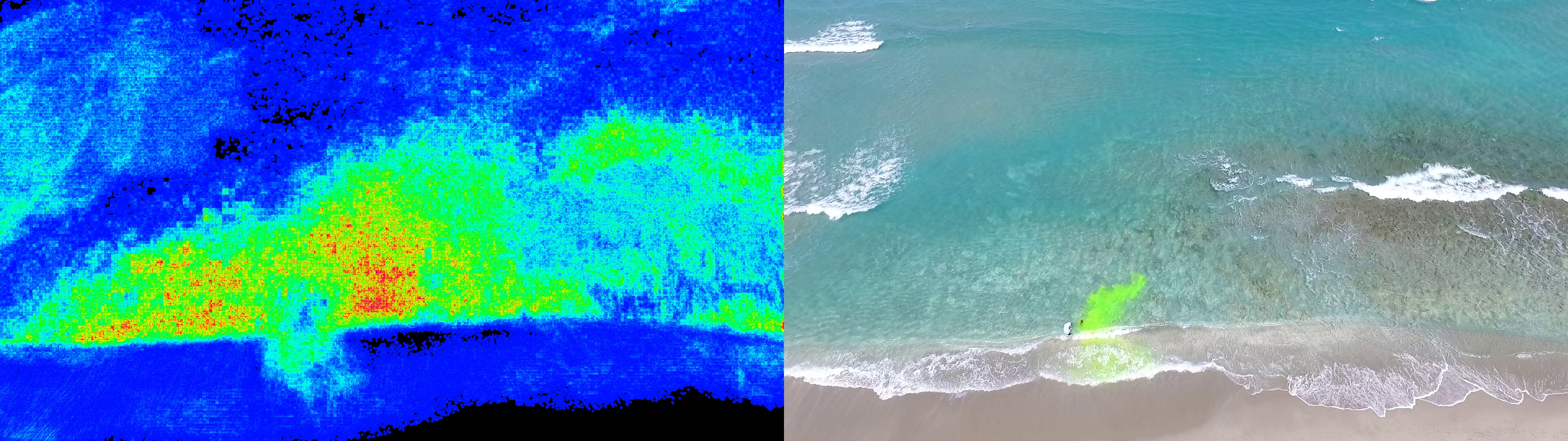}
	}
	\\
	\vspace{-0.6em}
	\setcounter{subfigure}{0}
	\subfloat[Original video]{
		\includegraphics[width=0.14\textwidth,trim={17.8cm 0 0 0},clip]{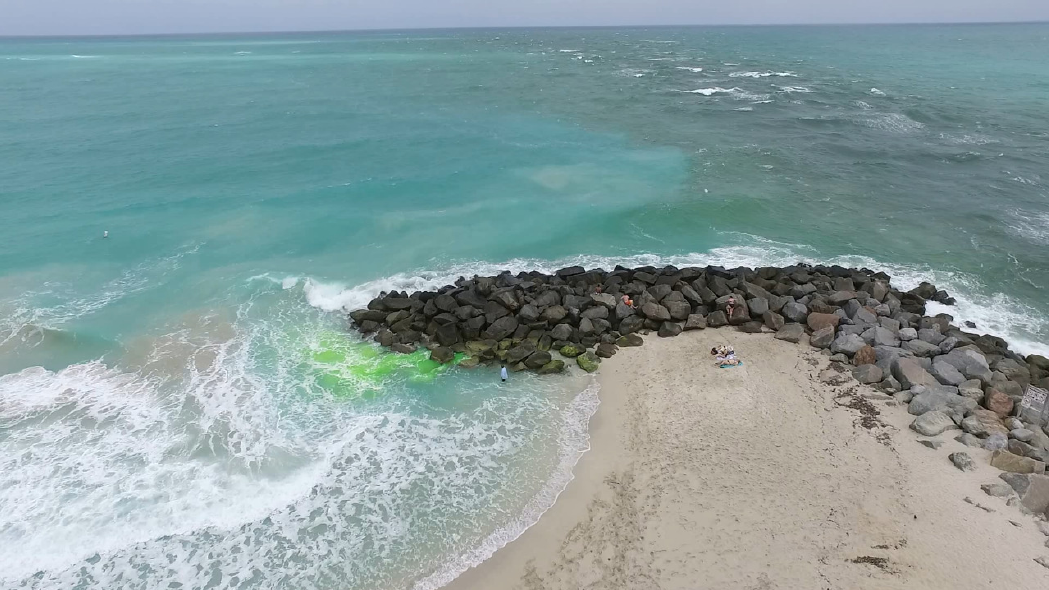}
	}
	\subfloat[LucasKanenda]{
		\includegraphics[width=0.14\textwidth,trim={0 0 67.7cm 0},clip]{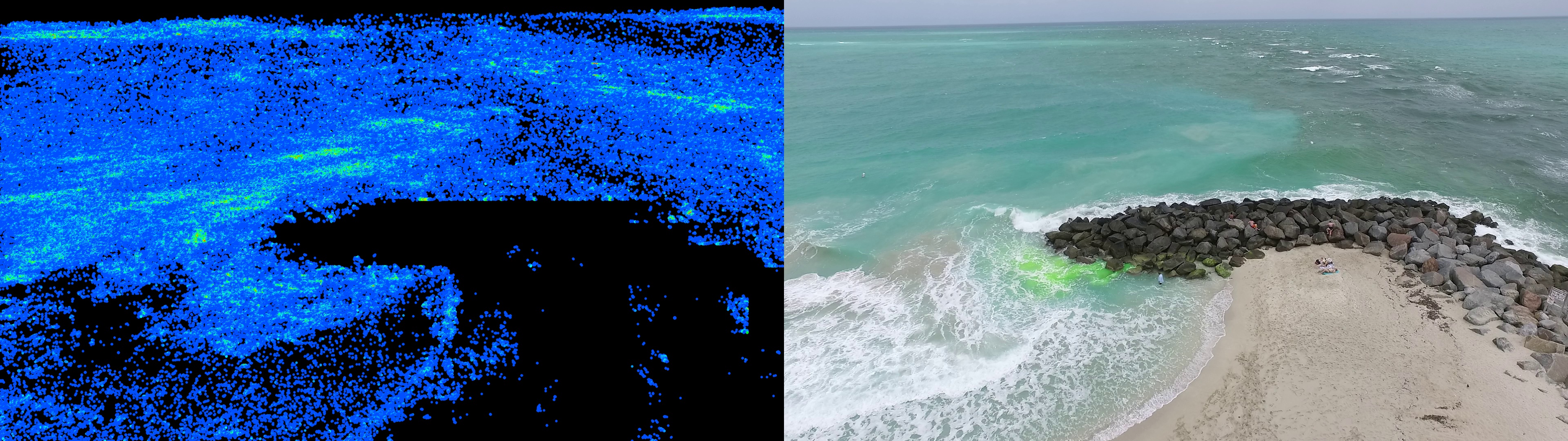}
	}
	\subfloat[HOR LucasKanenda]{
		\includegraphics[width=0.14\textwidth,trim={0 0 67.7cm 0},clip]{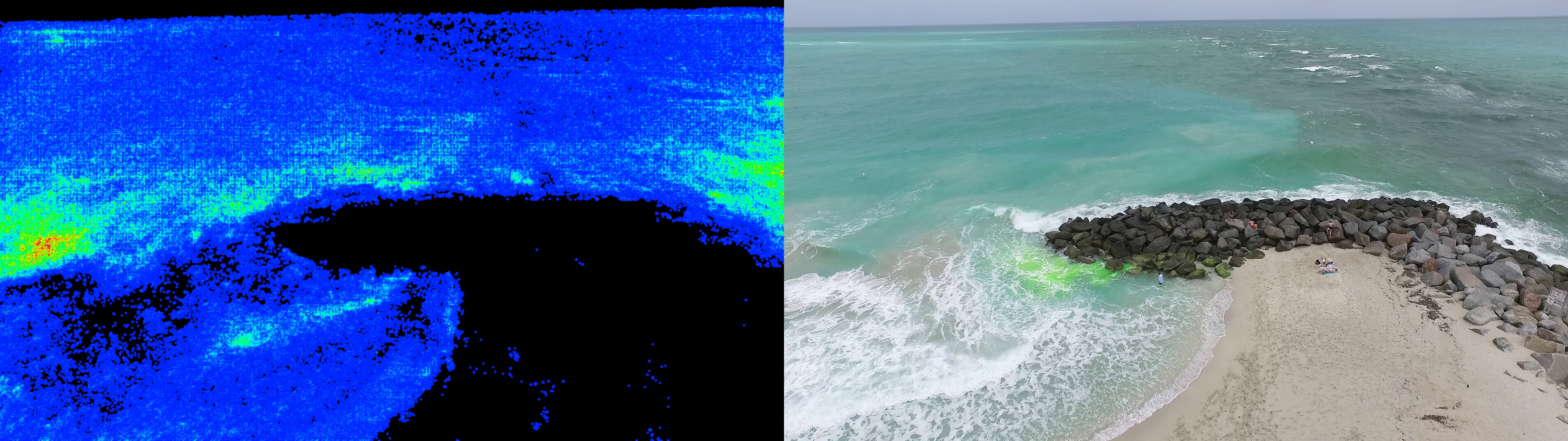}
	}
	\subfloat[HornSchunck]{
		\includegraphics[width=0.14\textwidth,trim={0 0 67.7cm 0},clip]{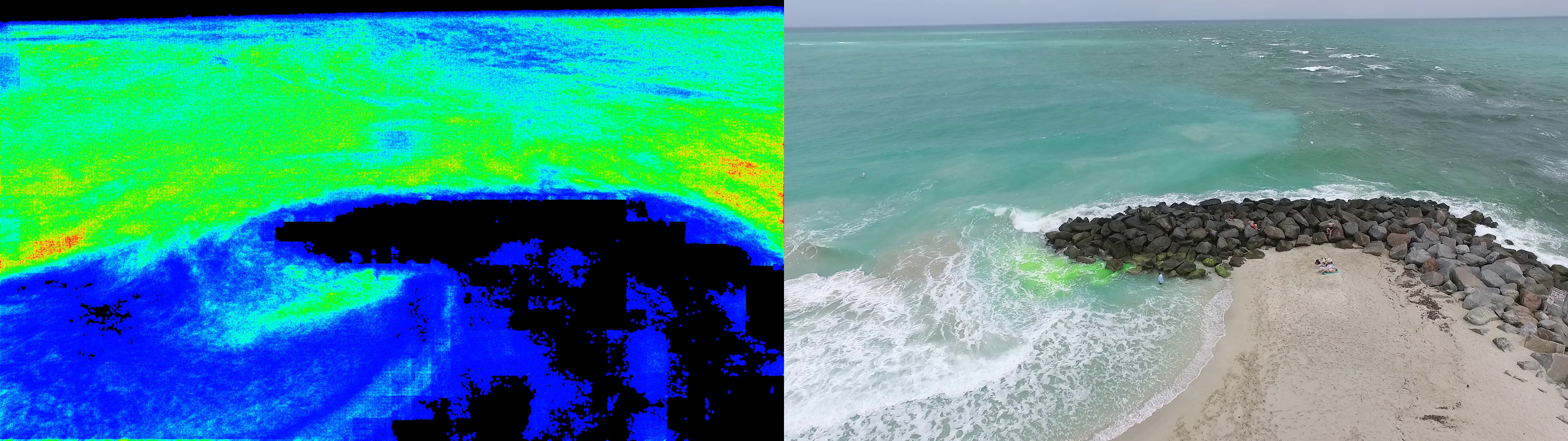}
	}
	\subfloat[HOR HornSchunck]{
		\includegraphics[width=0.14\textwidth,trim={0 0 67.7cm 0},clip]{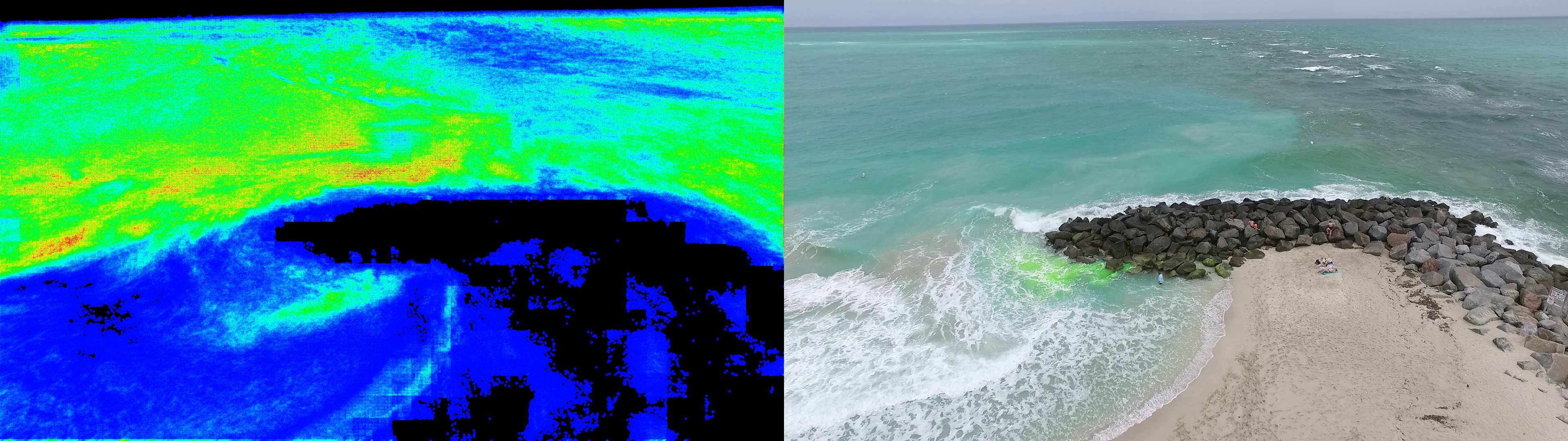}
	}
	\subfloat[Typhoon~\cite{typhoon}]{
		\includegraphics[width=0.14\textwidth,trim={0 0 67.7cm 0},clip]{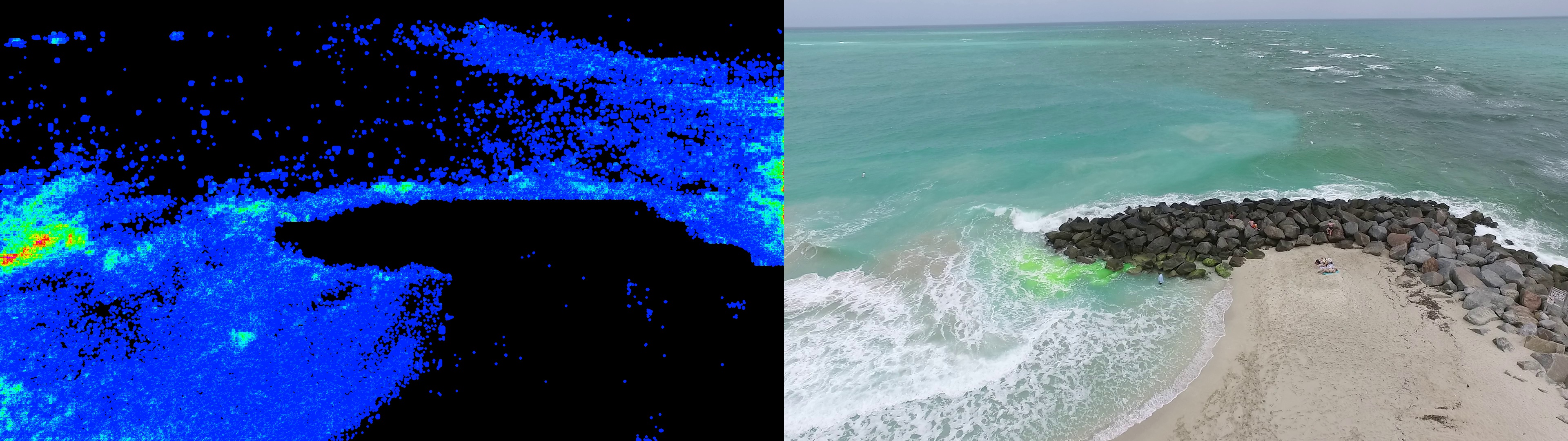}
	}
	\caption{Rip current likelihood heatmaps for the Palm Beach (in first row) and Haulover (in second row) videos using various optical flow estimation methods}
	\label{fig:heatmap}
	\vspace{-1em}
\end{figure*}

In rip current research, dye is commonly used to detect rip currents. To qualitatively evaluate the velocity estimation results of our proposed framework, we conducted a virtual drifter experiment. The experiment is designed based on the virtual drifter experiment used in~\cite{typhoon}. In the experiment, fluorescein (yellow) dye was released into the water, as shown in Fig.~\ref{fig:dye}.

We densely generated 400 virtual drifters in a $200 \times 200$ rectangular box with equal spacing and superimposed them on the original frame. The region of dye was thoroughly covered by the virtual drifters. As the velocity at each pixel can be obtained based on our proposed framework, we simulated the changes in virtual drifters' locations by the estimated velocity at the drifter locations. The positions of the virtual drifters were tracked for 20, 40, and 60 seconds, as shown in Fig.~\ref{fig:dye}(b)(c)(d). The results demonstrated that the virtual drifters followed the path of the rip current accurately despite outpacing the speed of the dye. The effectiveness of the estimated velocity generated by the optical flow methods was thus validated. However, it was also observed that the virtual drifters can be significantly affected by surface waves, indicating the importance of removing wave interference for rip current detection.

\begin{table}[]
	\centering
	\caption{The Evaluation Results of CNN-based Sea Water Segmentation}
	\label{tbl:seg_eval}
	\begin{tabular}{|c|c|c|}
		\hline
		\textbf{Video Name} & \textbf{IoU} & \textbf{F1-Score} \\ \hline
		Palm Beach      & 0.943 & 0.971 \\ \hline
		Haulover        & 0.828 & 0.919 \\ \hline
		Ocean Reef Park & 0.666 & 0.741 \\ \hline
		South Beach     & 0.935 & 0.969 \\ \hline
		\textbf{Overall} & \textbf{0.840} & \textbf{0.889} \\ \hline
	\end{tabular}
	\vspace{-1em}
\end{table}

\subsection{Sea Water Segmentation Performance Evaluation}
To train a U-Net model for sea water segmentation, we selected 233 frames from the four videos listed in Table~\ref{tab1}, which feature various scenes and objects. We split these frames into training and validation sets, with 203 frames used for training and the remaining 30 frames for validation. To achieve fine-grained segmentation results, we generated fifty $192 \times 216$ patches from each frame during training, with a stride of $96 \times 108$, resulting in a total of 10,150 training samples. During validation and testing, we generated 200 patches of the same size with a smaller stride of $48 \times 54$ as inputs, and we used a majority voting scheme to determine whether each pixel is sea water or not.

To evaluate the performance of our sea water segmentation model, we computed the Intersection over Union (IoU) and F1-Score metrics both overall and per-video on our validation set, and we present the results in Table~\ref{tbl:seg_eval}. Our model achieved excellent results in detecting sea water areas in video frames. We observed slightly lower performance in the Ocean Reef Park video, which featured more complex scenes with rare objects such as buildings. We believe that this issue can be addressed by expanding the training dataset with similar video frames.

\subsection{Rip Current Detection Performance Evaluation}
The proposed rip current detection framework estimates the probabilities of each pixel in the video having rip currents using the likelihood matrix. In Fig.~\ref{fig:heatmap}, we illustrate the results of both Haulover and Palm Beach videos based on various optical flow estimation methods. The heatmaps depict the likelihood of a rip current existing at a pixel, with higher probabilities shown in hotter colors. All results in this experiment use a time period $T=20$. Based on the results, we observe that Typhoon cannot detect most of the rip currents since it eliminates too many subtle currents in the video. The Lucas-Kaneda methods produce results with lower contrast against the background, leading to high false alarm rates and degraded performance. In contrast, the Horn-Schunch methods leverage global information for optical flow estimation and detect rip current regions more accurately, with the shape of high probability regions corresponding well to normal rip current regions. HOR (High-Order Regularization) further helps suppress noise in non-rip areas, resulting in better precision than both original Lucas-Kaneda and Horn-Schunch methods. Therefore, the HOR Horn-Schunch method that combines both advantages of global information and noise suppression achieves the best performance for rip current detection.

The Palm Beach video has a large oceanic area, with the sea taking up a significant portion of the frame. However, due to the rock outcropping, the right-middle part of the water is too shallow to be visible, resulting in potential noise from the seabed in this video.
Overall, we observe two high probability regions in the heatmaps. The lower-left region is a significant rip current area, certified by an established coastal science expert. Another high possibility region is located in the upper-right region.

\begin{figure}
	\centering
	\subfloat[Palm Beach video]{
		\includegraphics[width=0.14\textwidth]{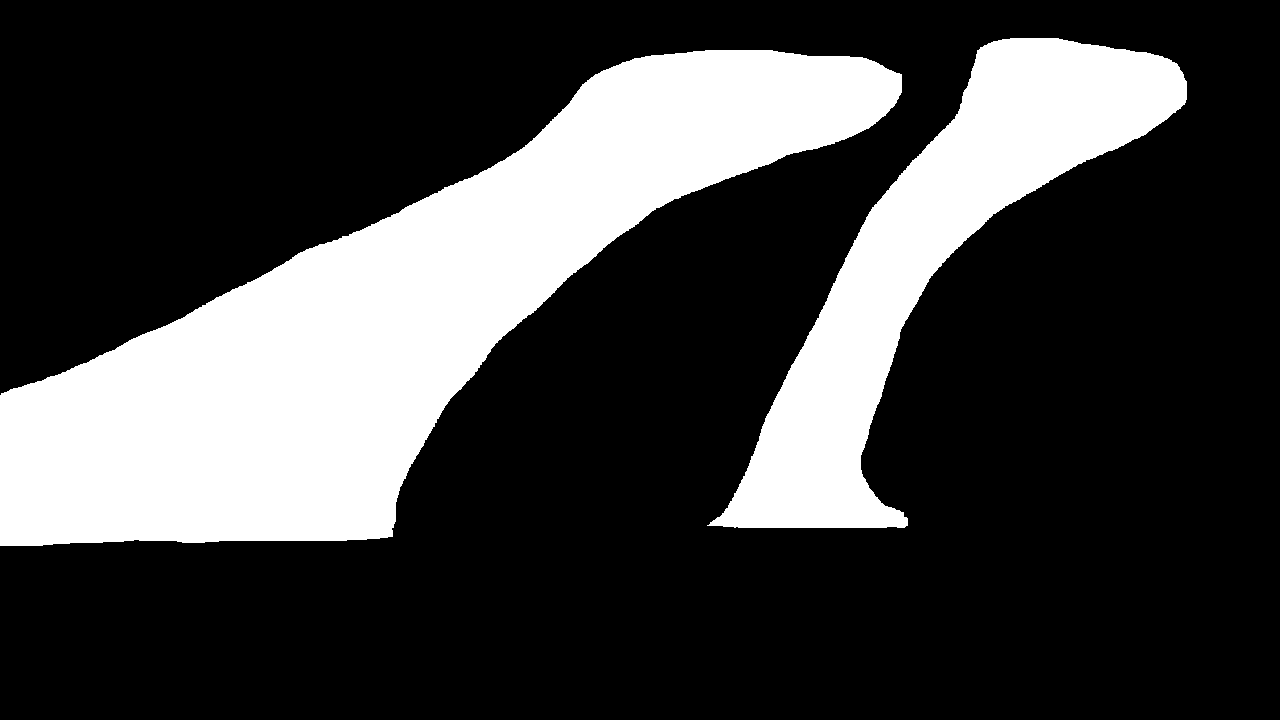}
	}
	\quad
	\subfloat[Haulover video]{
		\includegraphics[width=0.14\textwidth]{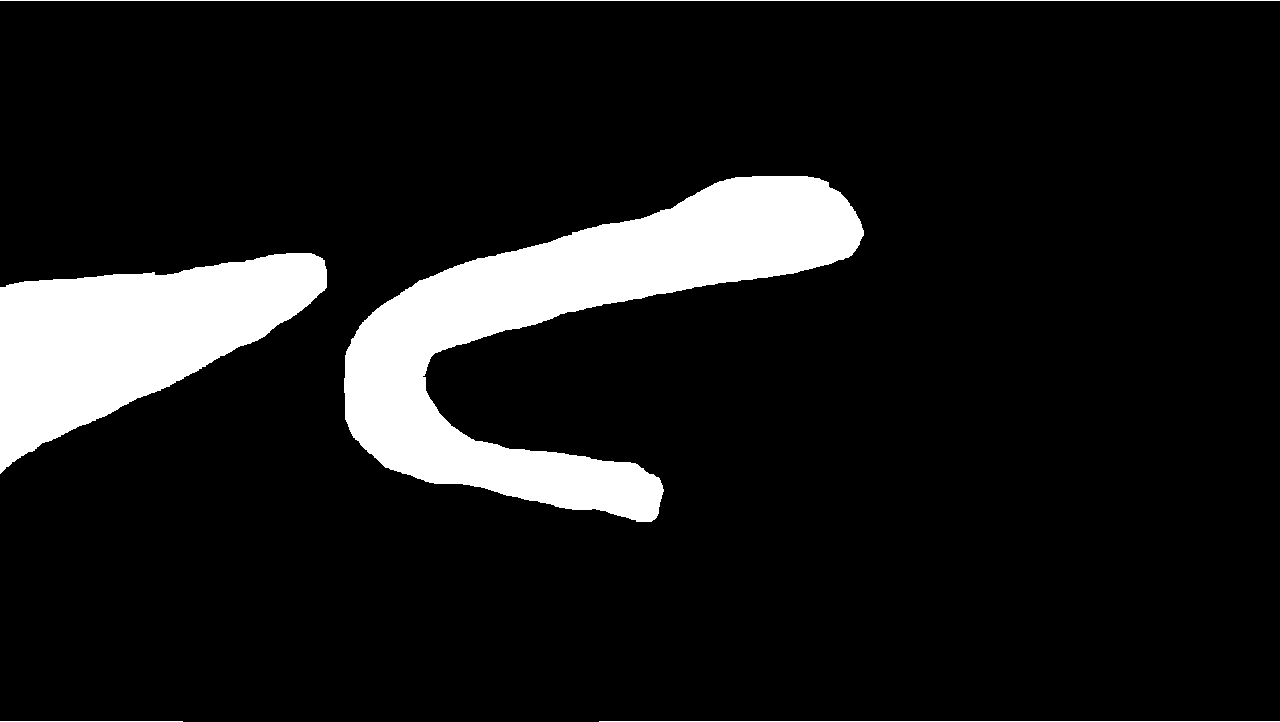}
	}
	\caption{Expert rip current region annotations for both videos}
	\label{fig:annotations}
	\vspace{-1em}
\end{figure}

\begin{figure}
	\centering
	\includegraphics[width=0.24\textwidth,trim={0 0 0 1.4cm},clip]{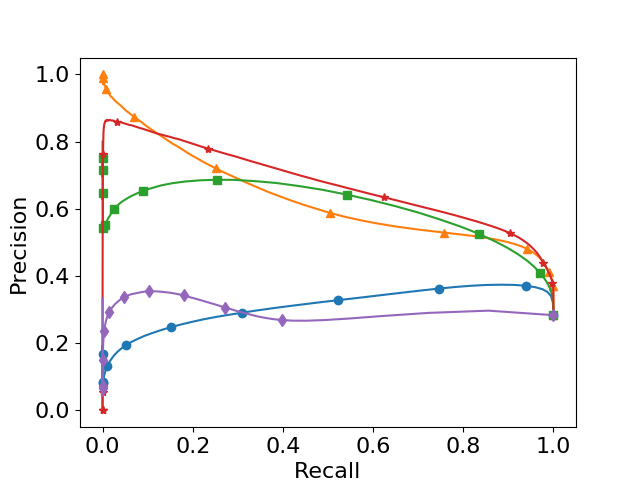}
	\includegraphics[width=0.24\textwidth,trim={0 0 0 1.4cm},clip]{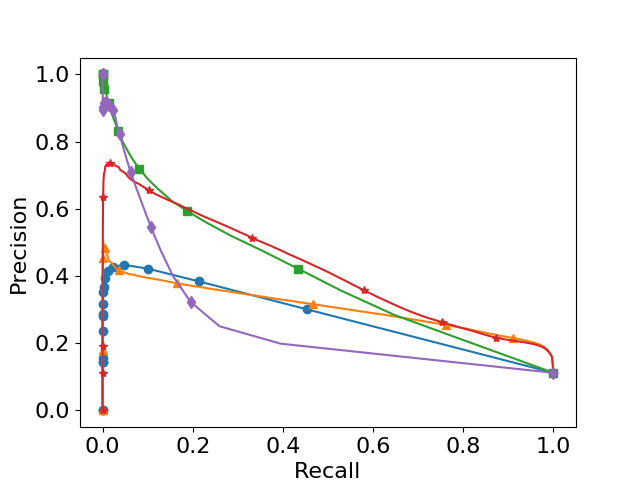}
	\caption{Comparison of various optical flow estimation methods ($\bullet$: LucasKanenda, $\blacktriangle$: HornSchunck, $\blacksquare$: HOR LucasKanenda, $\bigstar$: HOR HornSchunck, and $\blacklozenge$: Typhoon~\cite{typhoon}) for Palm Beach (left) and Haulover (right) videos}
	\label{fig:prc}
	\vspace{-1em}
\end{figure}

In this study, we evaluate the effectiveness of our proposed rip current detection framework by comparing its results with the ground truth rip current regions provided by three domain experts. The experts used their professional knowledge and experience based on features observed in the video frames, the fluorescein dye, and GPS drifters~\cite{leatherman2017rip, leatherman2017techniques} to annotate the rip current regions, as shown in Fig.\ref{fig:annotations}. To quantitatively assess the performance of our framework, we generate precision-recall curves, as shown in Fig.\ref{fig:prc}, which plot the values of precision and recall metrics at various levels of acceptance.

We define the region of rip current at a given level of acceptance $a$ as $R^a={(i,j)|L_{T,i,j}\geq a}$, where $L_{T,i,j}$ represents the likelihood matrix estimated by our framework for the pixel at $(i,j)$ and $a$ is the threshold. The precision and recall values at the pixel-level are then computed as follows:
\begin{equation}
	\text{precision}^a=\frac{|R^a\cap R_{gt}|}{|R^a|}, \text{recall}^a=\frac{|R^a\cap R_{gt}|}{|R_{gt}|}
\end{equation}
where $|\cdot|$ refers to the number of pixels in the given region and $R_{gt}$ is the region of rip current based on the expert's annotation.

We observe that the precision-recall curves for the different optical flow estimation methods exhibit distinct patterns. The precision typically increases as the recall decreases, indicating that more accurate results can be obtained when accepting fewer rip current regions. However, if the precision decreases instead, it suggests that some noise has been detected as rip currents with high probabilities. For instance, the Lucas-Kaneda method suffers from noise in areas with rocky seabeds, as seen in the right part of the Palm Beach video. Similarly, the Horn-Schunck method exhibits the same issue in the Haulover video, resulting in much lower precision scores than for the Palm Beach video, especially for the range of recall $<$ 0.5. As the HOR method effectively suppresses noise and demonstrates better robustness, its precision-recall curves exhibit similar patterns for both videos.

\begin{table}
	\centering
	\caption{The comparison of AUC for both videos}
	\label{tbl2}
	\begin{tabular}{|c|c|c|}
		\hline
		\multicolumn{1}{|l|}{\textbf{Optical Flow Estimation Method}} & \multicolumn{1}{l|}{\textbf{Palm Beach}} & \multicolumn{1}{l|}{\textbf{Haulover}} \\ \hline
		Typhoon~\cite{typhoon}                                                       &                         0.300                 &                   0.260                     \\ \hline
		Lucas-Kanenda                                     &                   0.316                       &                          0.283              \\ \hline
		HOR Lucas-Kanenda                                    &                   0.610                      &                               0.405         \\ \hline
		Horn-Schunck                                      &                 0.632                         &                               0.309         \\ \hline
		\textbf{HOR Horn-Schunck}                                     &              \textbf{0.677}                            &                   \textbf{0.421}                     \\ \hline
	\end{tabular}
	\vspace{-1em}
\end{table}

Moreover, the computation of the area under the precision-recall curve (AUC) is performed, which represents the area bounded by the precision-recall curve, above the line $\text{precision}=0$, to the right of the line $\text{recall}=0$, and to the left of the line $\text{recall}=1$. The AUC metrics for both videos based on all the optical flow estimation methods are presented in Table~\ref{tbl2}. The results demonstrate that the HOR method consistently outperforms the basic optical flow methods, and the Horn-Schunck method consistently performs better than the other methods.

Furthermore, the proposed framework using the HOR Horn-Schunck optical flow estimation method consistently achieves the best results for both videos, with AUC values of 0.677 for the Palm Beach video and 0.421 for the Haulover video. However, it should be noted that in the right part of the video, the detected rip current region is separated into two streams, whereas the expert annotation covers the entire area as one. Since there is no dye released in this area, it is impossible to validate the real pattern of the rip current regions. Nevertheless, the rip current has been clearly identified and matched with the expert's annotation.

In addition, the rip currents in the middle-bottom part of the Haulover video have almost reverse direction compared to the others, which violates the basic assumption of Eq.~\ref{eq:direction}. As a result, the proposed framework might fail to detect rip currents when there is no constant offshore direction in the video. In the case of a complex coastline, the proposed framework is limited in its ability to identify the correct offshore direction and may thus produce incorrect rip current detection results.

\section{Conclusion and Further Work}\label{sec:conclusion}
This paper proposes a novel framework for detecting rip currents based on UAV videos, which combines optical flow estimation and temporal data fusion. The proposed framework is evaluated using videos from two locations, and both qualitative and quantitative evaluations demonstrate its effectiveness in identifying rip current regions. The HOR Horn-Schunck algorithm achieves the best performance among all optical flow estimation methods, as it can suppress noise in the estimated velocity and capture the velocity of subtle currents in the videos.

Future work will focus on collecting more UAV videos of rip currents under different weather conditions, and extending the proposed framework for videos in rainy, foggy, and stormy weathers. In addition, the use of CNN for wave segmentation will be investigated to improve the performance further. To meet the need for real-time computation in real-world applications, low-power GPU will be used to accelerate the computations, and the system will be deployed on UAVs for in-field detection.

\section{Acknowledgments}\label{sec:ack}
This article is an extension of the master's thesis by Anchen Sun in 2020~\cite{asun2020}. We would like to thank S. B. Leatherman and S. P. Leatherman from the Department of Earth \& Environment, Florida International University, Miami, FL, USA, for generously providing us with the UAV videos and ground truth rip current regions for each video, which were essential to the success of this research project. We would also like to express our appreciation to Yudong Tao from Meta for his valuable suggestions on the writing style and organization of this article. Finally, we gratefully acknowledge the guidance and support of Dr. Mei-ling Shyu and Dr. Shu-Ching Chen throughout this research.

\appendices

\ifCLASSOPTIONcaptionsoff
  \newpage
\fi



\bibliographystyle{IEEEtran}
\bibliography{IEEEabrv,ieee_tgrs}
%



\begin{IEEEbiography}[{\includegraphics[width=1in,height=1.25in,clip,keepaspectratio]{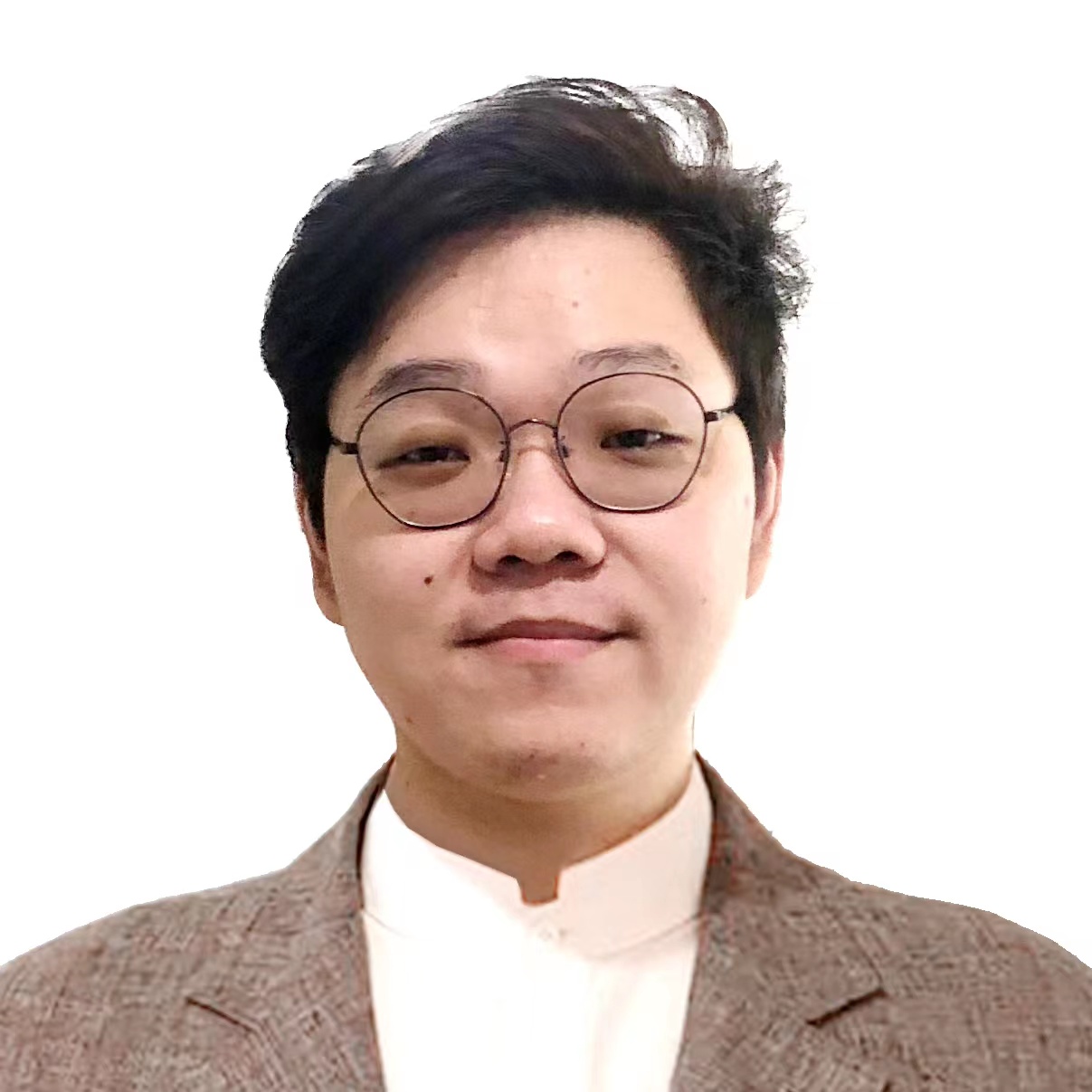}}]
{Anchen Sun} is a current Ph.D. candidate at the Department of ECE at UM, since August 2020. He received his B.S. degree in Marine and Atmosphere Science/Computer Science in 2017 and M.S. degree in Electrical and Computer Engineering in 2020 from UM, Coral Gables, FL, USA. He was IDSC Fellow in 2016-2017. His research interests include Deep Learning, Bioinformatics, Artificial Intelligence, Engineering Management.
\end{IEEEbiography}



\begin{IEEEbiography}[{\includegraphics[width=1in,height=1.25in,clip,keepaspectratio]{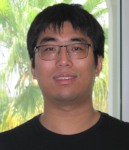}}]
{Kaiqi Yang} is an IMSA Research Assistant Professor at the Department of Mathematics at UM, since August 2023. He received his B.S. degree in Mathematics in 2016 from Hong Kong University of Science and Technology. Then he awarded his M.S. and Ph.D. degree in Mathematics in 2019 and 2024 from New York University, New York, NY, USA. His research interests include Algebraic Geometry.
\end{IEEEbiography}

\end{document}